\documentclass[conference,compsoc]{IEEEtran}

\usepackage{listings}

\ifCLASSOPTIONcompsoc
  \usepackage[nocompress]{cite}
\else
  \usepackage{cite}
\fi

\usepackage{amsfonts}  

\usepackage{amsmath}

\usepackage{hyperref}

\usepackage{graphicx}

\begin{document}
\bstctlcite{IEEEexample:BSTcontrol}

%
\title{Requirements for Ethereum Private Sidechains}


\author{
    \IEEEauthorblockN{Peter Robinson}
    \IEEEauthorblockA{Protocol Engineering Group and Systems, ConsenSys\\
    peter.robinson@consensys.net}
    \IEEEauthorblockA{School of Information Technology and Electrical Engineering, University of Queensland, Australia\\
    peter.robinson@uqconnect.edu.au}
}

\maketitle

\thispagestyle{plain}
\pagestyle{plain}

\begin{abstract}
The Enterprise Ethereum Client Specification by the Enterprise Ethereum Alliance defines the requirements which Ethereum Clients offering private smart contract capabilities should comply with. This specification though ground breaking, misses some important blockchain requirements and does not fully consider the requirements of Ethereum Clients offering Private Sidechain capabilities. This paper presents the case for Private Sidechains and defines requirements to be complied with to deliver this technology.

The capabilities of three blockchain clients have been analysed based on the requirements: Quorum, Parity, and Hyperledger Fabric. Quorum and Hyperledger Fabric operate as private consortium blockchains where as Parity delivers private transaction capabilities on top of Ethereum MainNet. These differing approaches has led to different strengths and weaknesses which has resulted in each client not complying with one or more key requirement. In particular, none of the reviewed blockchain clients support the ability to determine bootstrap information to establish on-demand blockchains and none of the clients support secure management and pinning from Ethereum MainNet. 

This paper presents Ethereum Private Sidechains and a range of technologies which allow it to deliver on complex sidechain requirements. Ethereum Registration Authorities are presented, which allow entities which have not previously interacted to securely obtain information to bootstrap a sidechain, and a Management and Pinning strategy is described which allows the state of a sidechain to be securely pinned to Ethereum MainNet without compromising privacy.

\end{abstract}


%
\IEEEpeerreviewmaketitle

\section{Introduction}
Ethereum \cite{wood2016a} is a blockchain platform which allows for the upload and execution of computer programs known as ``smart contracts". In the Ethereum public network, ``MainNet", all contract code and data are readable by any user of any node which connects to the network. Permissioning is a feature of systems in which authorised users are authenticated to use objects according to access control rules. In the context of Ethereum, permissioning could be used to limit which nodes connect to other nodes on the network, which users can use a node's JSON RPC API, and which Ethereum Accounts can use a smart contract. Ethereum MainNet can only perform permissioning in contract code, limiting which accounts can update the state of a contract. However, there is no mechanism to limit which users can read contract code and data. 

Enterprises need security and permissioning over and above what is available in standard Ethereum \cite{enteth10}. This has led to a range of platforms being developed to address the need for ``private smart contracts". J.P. Morgan developed Quorum \cite{quorum-source}, a fork of the Golang Ethereum implementation called, geth  \cite{geth-github}. Parity Technologies added a Private Transactions feature to their existing Ethereum client. Hyperledger Fabric \cite{androulaki2018} is a distributed ledger platform originally created by IBM and now hosted by The Linux Foundation. This platform directly competes with Quorum and Parity, offering privacy and permissioning features. Whereas Quorum and Parity offer private smart contracts which operate on or in conjunction with a permissionless blockchain, Hyperledger Fabric offers the ability to host one or more smart contracts on a private blockchain called a ``channel". Hyperledger Fabric allows multiple channels to be operated on the one network, thus allowing for multiple sets of private contracts between different sets of participants to operate on the one network.

Quorum and Hyperledger Fabric are private blockchain systems which are targeted at consortiums who wish to operate their own private networks. Parity offers its private contract feature on the public Ethereum MainNet. The paper proposes a middle ground, in which private permissioned blockchains can use the strong non-repudiation security guarantees of Ethereum MainNet.

The Ethereum community has been investigating how to scale the platform. Sharding \cite{sharding-definition} \cite{sharding-definition-uo}, a term borrowed from multi-user online gaming to describe how users are split across servers, and used by the database community to describe how data is split across servers, has been used in the Ethereum community to describe how the blockchain could be split such that smart contracts could operate on a partition of the permissionless public blockchain \cite{ethereum-sharding}. Similarly, Plasma \cite{poon2017} offers the possibility of operating multiple permissionless delegate public blockchains. 

Inspired by the concepts of Ethereum scaling, the concepts from Hyperledger Fabric of offering multiple permissioned channels, and of the privacy and permissioning features of Quorum, this paper proposes the concept of Ethereum Private Sidechains: permissioned Ethereum blockchains which host private smart contracts. Ethereum Private Sidechains offer permissioning such that only nodes which belong to organisations which are a party the sidechain have access to the sidechain transactions or state. The Ethereum Sidechain Client APIs are permissioned such that only authorised users can interact with them, and only authorised Ethereum Accounts can be used to submit transactions from them. The transactions and the state of sidechains are authenticated-encrypted to ensure data confidentiality. 

A contribution this paper proposes the concept of Ethereum Registration Authorities (ERAs). These smart contracts allow Ethereum Private Sidechains to be established on-demand, in a similar way to how a user of a web browser can establish a secure connection to a web server by simply entering in a URL such as \url{https://example.com/}. These smart contracts host look-up tables between domain names and other organisation based smart contracts. These organisation based smart contracts allow organisations to securely publish data, such as the mapping between domain names, IP addresses, and cryptographic keys. By being deployed on Ethereum MainNet, they are both public and secure. By being operated by reputable companies which undertake Know Your Customer \cite{know-your-customer} due diligence, and by operating the contracts on Ethereum MainNet which provides strong non-repudiation guarantees, users of ERAs will be safe in the knowledge that the data provided in these contracts can be trusted. 

A contribution this paper proposes a strategy for secure Management and Pinning, to allow the sidechain state to be pinned to a management chain. It outlines inter-sidechain requirements which will allow communications across chains.

This analysis is a snapshot in time. The implementations analysed are constantly releasing updates with new features. The Enterprise Ethereum Alliance are likely to release an updated edition of their specification.  It should be expected that the compliance of implementations with the requirements shown here will change with time, and the requirements that the implementations need to meet will change. However, this paper provides a template for how such analysis of future requirements and future software versions can be undertaken. 

The analysis undertaken in this paper has been done in good faith. Source code, example code, and documentation do not always stay in sync. Determination of conformance to many of the requirements is subjective. Please contact the author if you have questions or concerns about this paper.

Many terms are used inconsistently and at times unintentionally misleadingly in the blockchain field. To address this issue, terms used in this paper are listed in a glossary in Appendix \ref{glossary}. In particular, the term ``public" is used in Quorum to mean a permissionless blockchain. In Parity the term ``public" can mean Ethereum MainNet or it can mean a permissionless consortium blockchain. As such, throughout this document, a ``public" blockchain is a permissionless blockchain and a ``private" blockchain, sidechain or smart contract is a permissioned blockchain, sidechain or smart contract.

\subsection{Contributions}
The contributions made by this paper are:

\begin{itemize}  
\item Definition of additional private blockchain requirements not included in the Enterprise Ethereum Client Specification 1.0. These requirements have identifiers which are prefixed with $BC$.
\item Definition of sidechain specific requirements. These requirement have identifiers which are prefixed with $SC$.
\item Analysis of how Quorum, Parity, and Hyperledger Fabric solutions meet the Enterprise Ethereum Client Specification 1.0 requirements and the requirements laid out in this paper.
\item Presentation of a blockchain architecture and Enterprise Ethereum Client architecture which meets the needs of Ethereum Private Sidechain systems.
\item Description of the Ethereum Registration Authority system which allows for information to securely placed onto the Ethereum blockchain such that it is discoverable.
\item Description of the Management and Pinning contract which pins the state of a sidechain to Ethereum MainNet without revealing the identities of sidechain participants and whilst protecting the transaction rate of the sidechain.
\end{itemize}

\subsection{Organisation of Paper}
This paper starts by defining the requirements for Ethereum Private Sidechains. An example is presented to allow readers to more readily understand how sidechains could work. The  Enterprise Ethereum Client Specification v1.0 \cite{enteth10} is used as a basis for requirements. The requirements relating to blockchains in general that were missed during the development of the Enterprise Ethereum Client Specification are laid out. This is followed by the sidechain specific requirements. Quorum, Parity, and Hyperledger Fabric are then analysed based on the requirements. Doing this comparison helps draw out where existing solutions are deficient, and what needs to be done to deliver a better solution. A proposed architecture for Ethereum Private Sidechains is then presented.

\section{Requirements}
The Enterprise Ethereum Client Specification 1.0 \cite{enteth10} defines the requirements for Ethereum Clients which offer permissioning features. This specification has been written based on Ethereum clients which are in existence in early 2018, and attempted to be forward looking. The specification has missed some requirements which pertain to all Enterprise Ethereum clients. Additionally, there are sidechain specific requirements which are not captured. To understand the requirements of Ethereum Private Sidechains, an example will be presented. Based on this example, the requirements over and above those set-out in the Enterprise Ethereum Client Specification for a client supporting Ethereum Private Sidechains are presented.

To help understand the base-line requirements of the Enterprise Ethereum Client Specification 1.0, the requirements from the specification are included in Appendix A.

\subsection{Example Use Case}
\begin{figure}
  \includegraphics[width=\linewidth]{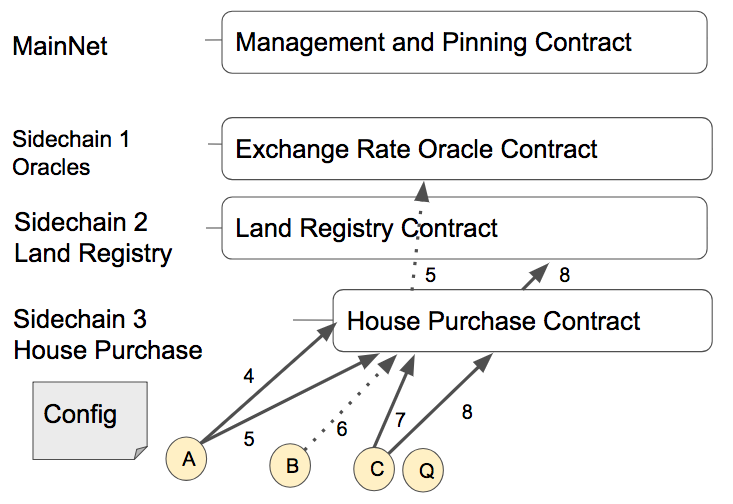}
  \caption{Usage Part 2}
  \label{fig:usage-part2}
\end{figure}

\begin{figure}
  \includegraphics[width=\linewidth]{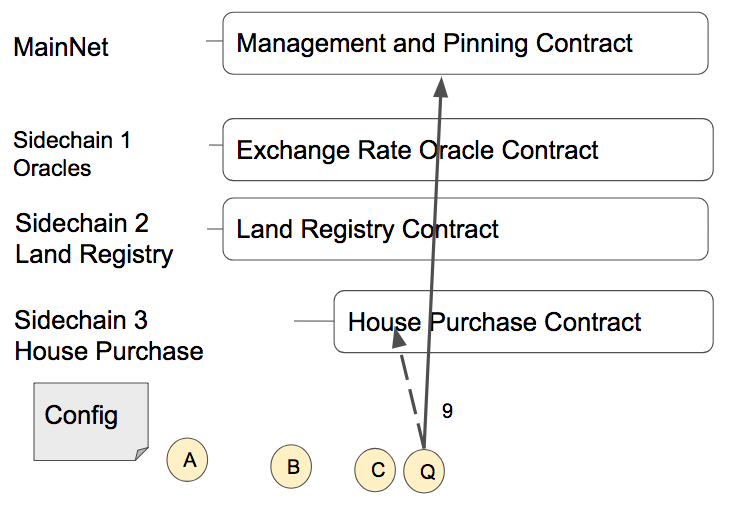}
  \caption{Usage Part 3}
  \label{fig:usage-part3}
\end{figure}

\begin{figure}
  \includegraphics[width=\linewidth]{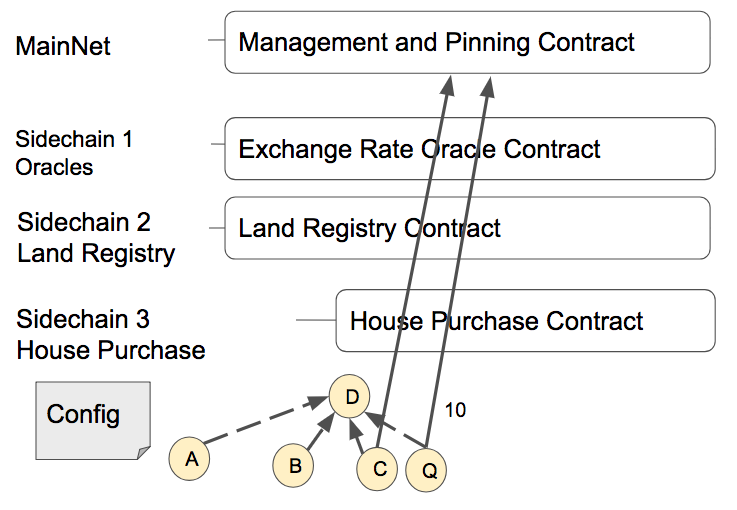}
  \caption{Usage Part 4}
  \label{fig:usage-part4}
\end{figure}

Figures ~\ref{fig:usage-part1} to ~\ref{fig:usage-part4} show an example usage of Ethereum Private Sidechains in which a house is purchased and then a land registry is updated based on the house purchase. Configuration and management of the sidechain is handled on Ethereum MainNet. Sidechain 1 holds an Oracle \cite{enteth10} contract from which data needs to be read. Sidechain 2 holds a land registry contract, which needs to be updated once the house purchase has been finalised.

\begin{figure}
  \includegraphics[width=\linewidth]{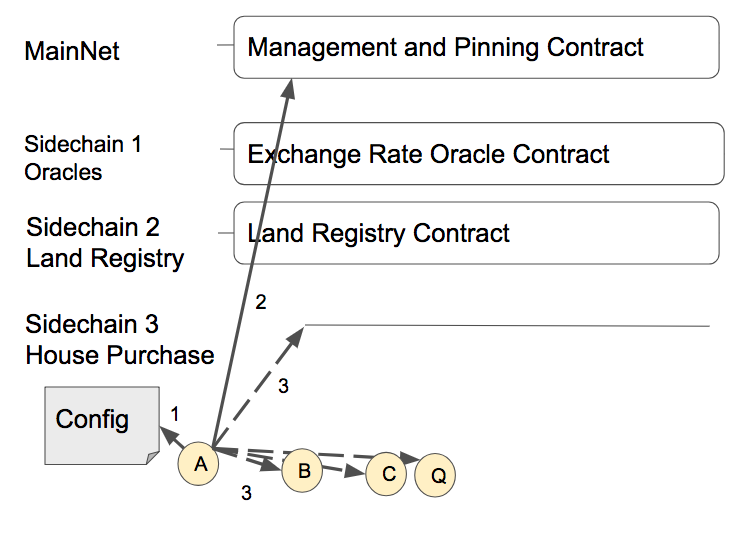}
  \caption{Usage Part 1}
  \label{fig:usage-part1}
\end{figure}

To complete the transaction data needs to be read from one sidechain and written to another sidechain. Walking through the example in detail:

\begin{enumerate}  
\item Participant A fetches configuration information from a Configuration entity. This information consists of addresses and keys and other information required to establish the sidechain (Figure ~\ref{fig:usage-part1}).
\item Participant A submits a transaction to register the sidechain with a Management and Pinning contract (Figure ~\ref{fig:usage-part1}).
\item Participant A creates the sidechain, inviting other entities to become participants of the sidechain (Figure ~\ref{fig:usage-part1}). 
\item Participant A submits a transaction which creates the House Purchase contract (Figure ~\ref{fig:usage-part2}).
\item Participant A submits a transaction, which results in fetching information from the Exchange Rate Oracle contract on Sidechain 1 (Figure ~\ref{fig:usage-part2}).
\item Participant B Fetches information from the House Purchase contract (Figure ~\ref{fig:usage-part2}).
\item Participant C submits a transaction to the House Purchase contract (Figure ~\ref{fig:usage-part2}).
\item Participant C submits a transaction in the House Purchase contract which causes a transaction to be submitted which updates the Land Registry contract (Figure ~\ref{fig:usage-part2}).
\item Participant Q could be a special node which only sees the encrypted state of Sidechain 3. It could read the state of the sidechain and post a representation of the state, known as a Pin, to the Management and Pinning contract. This pinning of the state could occur for each block or at regular intervals such as after every 10 blocks, or could be triggered to occur on demand (Figure ~\ref{fig:usage-part3}).
\item A new node can be added to the sidechain, gaining access to the House Purchase contract. The addition needs to be agreed upon using some technique. One technique could involve voting using the Management and Pinning contract (Figure ~\ref{fig:usage-part4}).
\end{enumerate}

\subsection{Additional Blockchain Requirements}
Derived from the example above, this paper proposes additional blockchain requirements over and above those listed in the Enterprise Ethereum Client Specification. 

\subsubsection{Permissioning and Credentials}
The Enterprise Ethereum Client Specification lists requirements related to authenticated participants that can execute JSON RPC API calls. The requirements do not differentiate between the different types of calls. For example, submitting transactions, performing read-only View calls  \cite{solidity}, or calling administrative APIs are all deemed the same. A user who is authenticated to use the API can execute any of the APIs. It could be imagined that in many cases having a node as a ``read-only" replica of the blockchain, from which applications could only read data but not execute transactions could improve security.

Additionally, there are no restrictions on which Ethereum accounts can be used to issue transactions from on a particular node. For example, if participant A in the example submitted transactions from a certain Ethereum Account A1, and participant B submitted transactions form an Account B1, it would be a security issue if an attacker, who has stolen participant A's private key could submit transactions from participant B's nodes. Adding controls to limit which Ethereum Accounts can be used with which node will improve the security of Enterprise Ethereum Clients.

Another situation which the Enterprise Ethereum Client Specification does not address is that different types of transactions should be treated differently. That is, an account which can submit a transaction to update the state of a contract should not necessarily be able to submit a transaction which deploys a new contract.

The requirements below specify the needs described in the paragraphs above:
\label{BC1c-TransactionTypePermissioning}
\begin{itemize}
\item BC-1a-ApiCallPermissioning: Implementations SHOULD provide separate access control permissions for API calls which are Administrative, API calls which allow viewing the state of the distributed ledger or are View calls, API calls which result in a transaction, API calls which result in deploying a contract, and API calls which transfer Ether.
\item BC-1b-EthereumAccountWhitelist: Implementations SHOULD allow for the specification of a whitelist of Ethereum Accounts which can be used to send transactions via the API. 
\item BC1c-TransactionTypePermissioning: Implementations SHOULD be able to authorise the types of transactions an Ethereum Account can submit. That is, separate permissioning should be provided for the ability to deploy a contract, call a function which changes the state of a contract, or performs a simple value transfer.
\end{itemize}

The Enterprise Ethereum Client Specification's requirement EE-6.1.2b-RestrictedPayloadMaskingStored and EE-6.1.2k-UnrestrictedPayloadMaskingStored relate to storing private transaction payloads securely. When these transactions are executed, they update the private state of the distribute ledger. Enterprise Ethereum Clients are likely to store the private state persistently, to reduce the restart time of a node. If the private state isn't stored, then all private transactions would need to be reprocessed each time an Enterprise Ethereum node restarted. As such, the follow requirement related to encryption of Enterprise Ethereum client's private state should be added to the Enterprise Ethereum Client Specification.
\label{BC-1d-PrivateStateAuthenticatedEncryption}
\begin{itemize}
\item BC-1d-PrivateStateAuthenticatedEncryption: If an implementation stores the private blockchain state persistently it MUST protect the data using an Authenticated Encryption with Additional Data (AEAD) algorithm such as one described in RFC5116, ``An Interface and Algorithms for Authenticated Encryption" \cite{rfc5116}.
\end{itemize}

\subsubsection{Organisations}
The Private Smart Contracts are likely to be between organisations. That is, in the example entities A, B, C, D, and Q may represent separate organisations. For reasons of systems resilience, organisations will want to operate at least four nodes. It can be envisaged that each organisation would operate nodes in at least two data centres. In this way, if one data centre becomes uncontactable due to network outages or power failure, the other data centre will continue to operate. Within each data centre, it can be envisaged that the organisation would have nodes in at least two racks of computers. In this way, if the power or network connectivity to one rack of computers fails, the data centre will still have an operational node. Computers which are connected to the nodes in the data centre will still be able to operate in a low latency environment, as they will still be able to contact a node inside the data centre. 

Organisations will host nodes on computers that will need to go offline and come online from time to time for reasons such as operating system updates and other software updates. Additionally, the organisation may decide to operate in a new data centre. In these situations, an Organisation needs to have a mechanism for adding and removing nodes to and from a Private Sidechain.

Some companies with many resources may operate dozens of nodes whereas a more resource constrained company with fewer resources may only operate a few nodes. All Ethereum Private Sidechain protocols need to be designed to be organisationally aware, and not give one company more power because it has more nodes that another company. 

Enterprise Ethereum networks generally, and Ethereum Private Sidechains in particular, are established between organisations who operate nodes. As such, it makes sense when establishing an Ethereum Private Sidechain to not denote the network to be set-up by Ethereum enode addresses of individual nodes, but by a higher level indication of the organisations involved. For example, domain names representing organisations could be used. Using human understandable notation such as domain names also allows for the possibility of establishing sidechains between organisations without prior arrangement. It allows the sidechains to be set-up without the need to exchange enode addresses and other configuration information. 

The Enterprise Ethereum Client Specification \cite{enteth10} includes requirement EE-5.1.1g-Organization, which states, ``An Enterprise Ethereum client SHOULD provide mechanisms to define clusters of nodes at the organisational level, in the context of permissioning." This requirement can be viewed as covering the need to address nodes using notation such as domain names, and providing the facility for an organisation to add nodes under its control to a sidechain as and when necessary. It however does not cover the need for protocols to be organisationally aware. 

\begin{itemize}
\item BC-2a-OrganisationallyAwareConsensus: Implementations SHOULD use organisationally aware consensus and voting algorithms.
\end{itemize}

\subsubsection{Bootstrapping}
Bootstrapping information needs to be available in a discoverable format. Doing this would allow sidechains to be created on-demand for previously unrelated parties. In the example, the configuration information could be in a discoverable format to allow participant A to find the enode address and encryption key information for the other entities. This then allows participant A to establish a sidechain immediately with the desired participants. This system of bootstrapping information for blockchain is analogous to the Domain Name System (DNS), Public Key Infrastructure (PKI), and Transport Layer Security (TLS) which combine to allow users of web browsers to enter a URL, for example \path{https://example.com/}, and be able to instantly, securely view information from a remote web server. Not having such a system of bootstrapping information for blockchain means that adding new participants to a network is a complex and time consuming process.

\begin{itemize}
\item BC-3a-DiscoverableBootstrapInfo: Implementations SHOULD provide a mechanism to discover bootstrap information for parties they have no previous relationship.
\end{itemize}

\subsubsection{Decentralisation}
A core tenant of Ethereum is decentralisation \cite{buterin2017a}. Architectural decentralisation refers to the property of a network in which there are no single points of failure. Political decentralisation refers to the property of a system where it is not controlled by a single or a small number of entities. Logical Decentralisation is the idea that a system may hold two values for the same item. For example, in the English language, the word ``Colour" is spelt with a ``u" in some countries and without a ``u" in others, whilst the meaning of the word remains the same. In blockchain systems, Architectural and Political Decentralisation is desirable, and Logical Decentralisation is not allowed as systems need to come to a consensus on the state of the distributed ledger. 

These concepts of Architectural and Political Decentralisation correspond to goals of the internet laid out some thirty years ago. Clark \cite{clark1988} when describing the design goals for the Internet refers to, ``Internet communication must continue despite loss of networks or gateways", as a primary goal and, ``The Internet architecture must permit distributed management of its resources", as one of the secondary goals. The fact that Architectural and Political Decentralisation were important when developing the internet provides a good indication of their importance in the blockchain era.

The following requirements relate to decentralisation.
\begin{itemize}
\item BC-4a-ArchitecturalDecentralisation: Implementations SHOULD not have any points of architectural centralisation.
\item BC-4b-PoliticalDecentralisation: Implementations SHOULD not have components which need to be controlled by a single entity or a small number of entities.
\end{itemize}

\subsubsection{Off-chain Messaging}
In the example, participant A could have been a company which represented an individual and participant B could have been a company which represents an individual. They may need to exchange the identity verification data they have about the individuals either with each other or they may need to distribute this information to all entities on the sidechain. In some jurisdictions, storing such sensitive information on a blockchain may not be desirable. As such, there is a need for Enterprise Ethereum solutions in general to offer off-chain communications capabilities. The off-chain communications method needs to ensure all communications is private. The requirements below reflect those needs.

\begin{itemize}
\item BC-5a-OffchainOrgToOrg: Implementations SHOULD provide an off-chain messaging mechanism which allows messages to be sent from one node to the nodes belonging to one or more organisations.
\item BC-5b-OffchainAll: Implementations SHOULD provide an off-chain messaging mechanism which allows messages to be sent from one node to the all nodes on the blockchain.
\item BC-5c-OffchianAntiSpam: Off-chain messaging mechanisms, if implemented, MUST have anti-spam features.
\item BC-5d-Whisper: Implementations SHOULD implement Whisper.
\end{itemize}

\subsubsection{Performance}
\label{EE-6.2.3b-ComputePowerSizeParticipants}The Enterprise Ethereum Client Specification requirements related to performance are not measurable and as such are not able to be met. Requirements EE-6.2.3b-ComputePowerSizeParticipants and EE-6.2.3c-RecentBlockAccessTime allow for infinitely large blockchain size by specifying, ``regardless of the blockchain size". Requirement EE-6.2.3b-ComputePowerSizeParticipants allows for an infinitely large number of participants by stating, ``regardless of the... number of network participants". These requirements need to be replaced with requirements which can be measured, are achievable, and meet customer expectations. The specification of such requirements is beyond the scope of this paper and is listed in the Future Work section.

\subsection{Additional Sidechain Requirements} Derived from the example above, this paper proposes requirements for sidechains over and above those listed in the Enterprise Ethereum Client Specification. 

\subsubsection{Permissioning and Credentials} Ethereum Private Sidechain implementations need to implement permissioning to restrict sidechain creation. In the example, participant A requested a node create a sidechain. Nodes under participant B, C, D, and Q's control accepted requests to join the sidechain. Permissioning also must be implemented to limit access between sidechains.  In the example, Sidechain 3 was able to perform a read-only View call \cite{solidity} from Sidechain 1 and perform a transaction on Sidechain 2.

It has been assumed that requirements EE-5.1.1cWhitelistNodes to EE-5.1.1f-BlacklistViaAPI apply on a per sidechain basis for Ethereum Private Sidechains. Thus the additional requirements required are as shown below.
\begin{itemize}
\item SC-1a-EstablishmentNodesWhitelist: It MUST be possible to specify a whitelist of node or organisation identities that are permitted to send sidechain establishment requests.
\item SC-1b-EstablishmentNodesBlacklist: Implementations MAY allow a blacklist of node or organisation identities to be specified that are not permitted to send sidechain establishment requests.
\end{itemize}

Only certain participant identities should be able to establish a sidechain. The requirements below reflect this need.
\begin{itemize}
\item SC-2a-EstablishmentApiWhitelist: Implementations MUST provide the ability to specify a white list of participant identities who are permitted to establish a sidechain.
\item SC-2b-EstablishmentApiBlacklist: Implementations MAY provide the ability to specify a black list of participant identities who are not permitted to establish a sidechain.
\end{itemize}

\subsubsection{Client Interfaces Sublayer} It is anticipated that most existing JSON-RPC API calls should have an additional parameter to indicate which sidechain they pertain to. Additional APIs need to be created which allow sidechains to be created on demand based on a list of domain names. As such the API requirements for sidechains are shown below.

\begin{itemize}
\item SC-3a-SidechainFindOrEstablishmentApi: Implementations MUST have an API which takes a set of domain names, organisational identifiers or node enode addresses, and requests the node find an existing sidechain, or create a new sidechain if no such existing sidechain exists. The function returns a sidechain identifier.
\item SC-3b-SidechainIdentifier: Implementations MUST allow an optional ``sidechain identifier" parameter to be specified for existing API calls. In the case of the    eth\_sendTransactionAsync function, an alternative function which takes a sidechain instead of a set of recipient public keys MUST be implemented.
\end{itemize}

Further analysis is needed to determine the specific administration APIs needed for adding and removing participants of a sidechain.

\subsubsection{Inter-Chain} EE-5.3.2a-InterChainInteraction \cite{enteth10} states, ``Enterprise Ethereum implementations MAY provide inter-chain mediation capabilities to enable interaction with different blockchains." This requirement relates to communication between blockchain platforms, such as between Quorum and Hyperledger Fabric, or between two instances of the same platform, or between two sidechains within the same instance of a platform. The word ``interaction" allows for many types of communications, including pinning the state of a sidechain or blockchain. More specific requirements are needed describe precisely what should be implemented to provide sidechain pinning capabilities.

As per the example, the state of Sidechain 3 needs to be able to be pinned to the management chain, MainNet. The pinning needs to be done in such as way that the list of participants on Sidechain 3 is not revealed on the MainNet as part of the pinning process. The rate of transactions need to be masked such that participants on MainNet can not infer the activity level on the sidechain. The frequency of pinning needs to be configurable. That is, pinning could be done for each block, or after a certain number of blocks, or at certain time intervals.

The pins of the sidechain state, once posted to MainNet must be able to be contested by members of the sidechain. It could be expected that the act of contesting a pin may require revealing of the identity on MainNet. 

It may be possible for special participants of the sidechain to only see cipher text. These parties could post pins of the encrypted state to MainNet, but not know the plaintext that the are providing the attestation for. Doing this means that sidechain participants who can see the plaintext of a channel do not have to reveal themselves on MainNet by the act of pinning. Participant Q in the example could be such a participant. 

The requirements below address these needs.

\begin{itemize}
\item SC-4a-Pinning: The state of the sidechain MUST be able to be pinned to a management chain or Ethereum MainNet.
\item SC-4b-PinningParticipantShielding: The list of participants of a sidechain SHOULD be able to be shielded when pins are posted to the management chain or Ethereum MainNet.
\item SC-4c-PinningTransactionRateShielding: The transaction rate of a sidechain SHOULD be able to be shielded when pins are posted to a management chain or Ethereum MainNet.
\item SC-4d-PinningContesting: Implementations MUST provide a mechanism for sidechain participants to contest a pin posted to the management chain. The act of contesting a pin may require the sidechain participant to reveal themselves.
\item SC-4e-PinningCipherTextObservers: Implementations MAY provide a mechanism such that special nodes that can observe the cipher text transactions on the sidechain, that do not have access to the plaintext transactions, can post pins of the sidechain state to the management chain or Ethereum MainNet.
\item SC-4f-PinningConfiguration: Implementations SHOULD allow cross chain pinning frequency to be configurable. That is, pinning could be done for each block, or after a certain number of blocks, or at certain time intervals.
\end{itemize}

For sidechain systems to be effective, Ethereum clients which support sidechains need to allow a multitude of sidechains to operate simultaneously. In the example there were three sidechain in addition to the management chain. Ethereum Private Sidechain clients need to have a specific requirement to support multiple clients.

\begin{itemize}
\item SC-4g-MultipleSidechains: Implementations MUST be able to operate multiple sidechains simultaneously.
\end{itemize}

\subsubsection{Privacy Sublayer}
Requirement EE-6.1.2r-PrivateTransactionAddParticipants states, ``Implementations SHOULD be able to extend the set of participants in a private transaction (or forward the private transaction in some way)." In the context of Ethereum Private Sidechains, adding a participant to a private transaction means adding an organisation and its nodes to a sidechain, or an organisation adding a node to a sidechain.

\subsubsection{Storage and Ledger Sublayer}
Requirement EE-7.1c-SeparateStoragePerNetwork states, ``Implementations providing support for multiple networks (for example, one or more consortium networks or a public network) MUST store data related to private transactions for those networks in private state dedicated to the relevant network." In the context of Ethereum Private Sidechains, this requirement should be interpreted to mean that the state of each sidechain should be encrypted with different keys and stored logically separately.

Requirement EE-7.1d-DataAccessSameParticipants \cite{enteth10}, ``A smart contract operating on private state SHOULD be permitted to access private state created by other smart contracts involving the same participants." can be interpreted as meaning smart contracts on a sidechain can access other smart contracts on the same sidechain. Requirement EE-7.1e-DataAccessDifferentParticipants, ``A smart contract operating on private state MUST NOT be permitted to access private state created by other smart contracts involving different participants." relates to a smart contract on one sidechain accessing a smart contract on another sidechain. In the example, the Exchange Rate Oracle Contract on Sidechain 1 could be operated by a company which provided access to the data of the Oracle for a fee. The participants who could submit transactions to update the state of the sidechain could be restricted to the operators of the sidechain. Any entity prepared to pay for access could become a participant which could read from the sidechain. As such, the participants of Sidechain 2 will be different to the participants of Sidechain 3. Additionally, in the example, Sidechain 2 would be operated by a government department tasked with operating the land registry. They might allow certain parties under certain circumstances to submit transactions to the land registry contract. The parties involved in such a contract will be different from the participants of the House Purchase Contract on sidechain 3. As such, to support Ethereum Private Sidechains, the requirement EE-7.1e-DataAccessDifferentParticipants should be removed and in their place there should be the following requirements.

\begin{itemize}
\item SC-5a-DataAccessDifferentParticipants: ``A smart contract operating on private state SHOULD be permitted to access private state created by other smart contracts which at involve an overlapping participant list, where the accessed private state resides on a chain which is configured to allow such access."
\end{itemize}

\subsubsection{Synchronisation and Disaster Recovery} In the example, after some period of time, the details of the House Purchase Contract and the entire contents of the sidechain may become irrelevant. To reduce costs of maintaining the data, the sidechain may need to be archived. The requirement below addresses this need.

\begin{itemize}
\item SC-6a-SidechainArchive: Implementations MAY provide the ability to archive a sidechain.
\end{itemize}

\section{Existing Privacy Solutions}
\subsection{Introduction}
The sections below analyse how existing distributed ledger privacy solutions meet the Enterprise Ethereum Client Specification requirements and the Ethereum Private Sidechain requirements described in this document. Two Ethereum based privacy technologies are analysed: J.P. Morgan's Quorum \cite{quorum-source} and Parity Technologies' Private Transactions \cite{parity-private-transactions}. Additionally, a non-Ethereum technology, Hyperledger Fabric \cite{hyperledger-fabric} \cite{androulaki2018} is analysed. Analysing Hyperledger Fabric against these requirements is important as it presents itself as a good solution for consortium networks.

Summary information of the analysis is presented in Tables ~\ref{table_comparison1} to ~\ref{table_comparison6} which are at the end of this paper. Section 6.1.4 of the Enterprise Ethereum Client Specification describes three Privacy Levels. Table ~\ref{table_comparison1} shows the Enterprise Ethereum Privacy Level C requirements. To be compliant with this level, all of the requirements in the Enterprise Ethereum Client Specification indicated as MUST or MUST NOT must be complied with; with the exception that implementations need to be compliant with only either the restricted private transactions or the unrestricted private transactions from section 6.1.2. Table ~\ref{table_comparison2} shows the Enterprise Ethereum Privacy Level B requirements. To be compliant with this level, in addition to the requirements for Privacy Level C, the implementation must be compliant with all of the requirements in sections 5.1.1, 5.1.2, and 5.1.3 marked as SHOULD in the Enterprise Ethereum Client Specification. Table ~\ref{table_comparison3} shows the Enterprise Ethereum Privacy Level A requirements. To be compliant with this level, in addition to complying with the requirements of Privacy Level B, the implementation must comply with all of the requirements indicated as SHOULD in section 6.1.2 of the Enterprise Ethereum Client Specification. Table ~\ref{table_comparison4} shows all other requirements described in the Enterprise Enterprise Ethereum Client Specification. Table ~\ref{table_comparison5} and ~\ref{table_comparison6} present novel requirements in addition to the Enterprise Ethereum Client Specification requirements. Table ~\ref{table_comparison5} shows the additional requirements which pertain to all Enterprise Ethereum blockchain clients. Ethereum Private Sidechain specific requirements are described in Table ~\ref{table_comparison6}.

In the tables, requirements which are complied with are indicated by $\checkmark$. Table cells containing a $\times$ indicates a requirement denoted as MUST, MUST NOT, SHOULD, or SHOULD NOT that is not complied with. Requirements which are denoted as MAY which are not implemented or not complied with are indicated in the tables with a dash (-).

In the tables, Ethereum specific requirements that Hyperledger Fabric does not meet as it is not an Ethereum based platform are marked as ``Ethereum". Ethereum specific requirements relate to forming consensus with Ethereum MainNet, using Ethereum JSON RPC APIs, or implementing specific Ethereum protocols such as DEVp2p and Whisper. The requirements which are deemed to be Ethereum specific are: EE-4.1a-DApp, EE-5.3.1a-JsonRpcPublicEth, EE-5.3.1b-JsonRpcTransactionAsyncExt, EE-5.3.1c-JsonRpcUnimplemented, EE-6.1.2h-RestrictedDefaultSecure, EE-7.1a-StoragePubEth, EE-7.2a-EvmOpCodes, EE-7.2b-EvmExtendedOpCodes, EE-7.2c-PublicStateSync, EE-7.2d-PrecompiledContracts, EE-7.3a-MainNetConsensus, EE-8.1b-DevP2P, EE-8.1c-Eth62Eth63, EE-8.1d-NewProtocols, EE-10a-PublicEthCompatibility, EE-10b-ExtendedApisSuperset, and BC-5d-Whisper. For requirements where analogies can be made between platforms, then these are deemed valid requirements.

\subsection{Quorum}
\subsubsection{Introduction}
JP Morgan created Quorum \cite{quorum-source}, a fork of geth \cite{geth-github} which provides Private Smart Contracts for Ethereum based consortium networks. Quorum operates a blockchain which communicates ``public" transactions and special ``private" transactions which just contain the message digest of the private transaction payload. Encrypted payloads are communicated between nodes using a private transaction manager component called ``Constellation". In addition to communicating transactions only to a specified set of nodes, Constellation is responsible for storage of the encrypted transactions for a node. The ``public" blockchain is similar to a typical Ethereum instance. It is referred to as ``public" as it is not permissioned, whereas the private state is permissioned. Both the public state and private state are stored in Quorum's database unencrypted.

To set-up a Quorum network, the enode addresses for each node are shared out of band with all nodes. Enode addresses combine an Ethereum address, an IP address, and UDP and TCP port for devP2P discovery and operational communications. To add or remove a node to the Quorum network, enode addresses must be supplied to all nodes.

Public transactions to the public blockchain are submitted using standard Ethereum JSON RPC calls. Private transactions must be submitted using the API similar but incompatible with the one described in EE-5.3.1b-JsonRpc2. The recipients of the private transaction are specified using the public encryption key of the destination nodes. 

Processing of private transactions is as follows \cite{quorum-wiki}: Quorum determines that the transaction is private by detecting the ``privateFor" field in the JSON RPC call and passes the transaction to Constellation. The transaction payload is encrypted using a randomly generated symmetric key. The message digest of the cipher text is kept as a digest of the encrypted transaction. The symmetric key of the encrypted transaction is encrypted against the public encryption keys for each recipient node specified in the ``privateFor" field. The encrypted transaction along with encrypted symmetric key and digest of the encrypted transaction are sent to each node participating in the transaction. The recipient node stores the encrypted symmetric key and encrypted transaction in a map using the digest of the encrypted transaction as the key. Once the encrypted payload has been sent to each Constellation node, a special public transaction which is marked as holding a private transaction payload digest is submitted to the public blockchain. The transaction is marked as being private by having a ``private" flag in the Ethereum transaction ``V" field. When the public transaction becomes part of a finalised block, each node checks its private transaction map, using the digest of the encrypted transaction as they key. If an encrypted transaction is found, then the transaction is decrypted and processed using the private state Ethereum Virtual Machine. 

Quorum offers the ability for Private Smart Contracts stored in the private state to call functions on Smart Contracts in the public state. The Private Smart Contracts can not alter the state of a Smart Contract stored in the public state, and as such should only be called on ``View" functions. 

The following sections analyse how Quorum meets the Enterprise Ethereum Client Specification requirements, the additional blockchain requirements, and the additional Sidechain requirements. The analysis has been based on Quorum's documentation \cite{quorum-wiki}, its examples \cite{quorum-examples}, source code \cite{quorum-source}, and conversations with people who have used Quorum in Proof of Concept deployments.

\subsubsection{Application Layer}
As Quorum extends the capabilities of geth, it allows users to make use of many of the tools which have been and continue to be created for use with geth. This means that many requirements are conformed with by virtue of Quorum extending geth. The paragraphs below analyse the Application Layer requirements.

EE-4.1a-DApp: Quorum does not support the JSON-RPC API Enterprise Ethereum extension described in EE-5.3.1b-JsonRpcTransactionAsyncExt, and as such does not comply with this requirement.

EE-4.3a-Tools: As Quorum extends geth, it is able to leverage the standard Ethereum smart contract deployment and debugging tools for use with its Enterprise Ethereum smart contracts. Quorum is compliant with this requirement.

EE-4.3b- FormalVerification: Any formal verification methods which become available for use with standard Ethereum are likely to work with Enterprise Ethereum contracts in Quorum. As such, Quorum should be deemed compliant with this requirement.

\subsubsection{Tooling Layer}
Quorum offers node permissioning but does not provide any authentication or access control of participants using its API. As such, where as many of the node based requirements are supported, most of the participant based requirements are not supported. The paragraphs below analyse each requirement.

EE-5.1.1a-StaticStartUp: Nodes which are a party to a Quorum network are specified in a file which needs to be securely shared with all participants out of band. As such, Quorum complies with the requirement to specify at startup a list of static peer nodes.

EE-5.1.1b-DisableDiscovery: Quorum does not allow for discovery of nodes in its Constellation network. As such, Quorum is not compliant with the requirement to be able to enable or disable peer-to-peer node discovery.

EE-5.1.1c-WhitelistNodes: Nodes which are to connect to the Quorum network must be explicitly enabled in a configuration file. As such, Quorum complies with the requirement to provide the ability to specify a whitelist of the node identities permitted to join the network.

EE-5.1.1d-BlacklistNodes: Quorum does not support blacklisting of nodes. 

EE-5.1.1e-WhitelistViaAPI: Quorum does not provide the ability to programmatically update the list of whitelisted nodes. As such, Quorum does not comply with this requirement.

EE-5.1.1f-BlacklistViaAPI: This requirement is conditional on the blacklisting support from requirement EE-5.1.1d-BlacklistNodes. As such, this requirement is not applicable. 

EE-5.1.1g-CertifyNodes: Nodes are identified by enode addresses. These addresses are shared out of band between all participants. There is no certification process as such for these identifiers. As such, Quorum does not comply with this requirement.

EE-5.1.1h-Organization: Quorum does not support a grouping or clustering of nodes for permissioning purposes. As such, Quorum does not comply with this requirement.

Quorum does not provide any mechanism to authenticate users of the JSON RPC API. As such, there is no ability to whitelist or blacklist participant identities who are permitted to submit transactions or use any other API. Quorum does not comply with requirements EE-5.1.2a-WhitelistParticipants, EE-5.1.2b-BlacklistParticipants, EE-5.1.2c-WhitelistParticipantsViaAPI, EE-5.1.2e-CertifyParticipants, and EE-5.1.2f-GroupsRoles. The requirement EE-5.1.2d-BlacklistParticipantsViaAPI is conditional on the blacklisting support from requirement EE-5.1.2b-BlacklistParticipants. As such, this requirement is not applicable. 

EE-5.1.3a-SmartContractPermissioning: Quorum allows contracts written in any Ethereum Virtual Machine language. This allows application developers to write their contracts such that they use address based permissioning in their smart contracts. This allows Quorum to comply with this requirement.

EE-5.1.3b-RuntimeConfigUpdate: Quorum allows the configuration to be updated such that new nodes are added to Quorum networks. The configuration can also be updated to remove nodes, however these nodes will not be removed unless the network connection between nodes is disrupted or the node restarts. As such, Quorum partially complies with this requirement.

EE-5.1.3c-ConfigOptions: Quorum provides configuration via flat files. As such, Quorum complies with this requirement.

EE-5.1.3d-LocalKeyManagement: Quorum offers a simplistic key management approach in which key pairs are generated during an initialisation step and then stored in a file. No key roll-over or revocation is possible. Despite the Quorum's key management approach limitations, it does comply with this requirement as it does support local key management to allow users to secure their private keys.

EE-5.1.3e-SecureExternalKeyGenStore: Quorum does not support external key generation or key storage, and as such does not comply with this requirement.

EE-5.1.3f-HardwareSecurityModules: Quorum does not support Hardware Security Modules and as such does not comply with this requirement.

EE-5.2.1a-IntegrationLibraries: There are a variety of integration libraries available for interacting with Quorum. For example, web3j supports Java integration with Quorum \cite{web3j-github}.  

Quorum does not support enterprise software deployment and configuration systems, software fault reporting capabilities, performance management capabilities, security management interaction capabilities, historical analysis, or enterprise management systems. However, products such as Kaleido \cite{kaleido} which provide Quorum as a managed service, allow users of Quorum to have enterprise capabilities. As such, requirements EE-5.2.2a-EntDeployment, EE-5.2.2b-EntFaultReporting, EE-5.2.2c-EntPerformanceManage, EE-5.2.2d-EntSecurity, and EE-5.2.2e-EntHistoricalAnalysis are marked as compliant when Quorum is used in conjunction with Kaleido.

EE-5.3.1a-JsonRpcPublicEth: Quorum complies with this requirement as it supports the public Ethereum JSON-RPC API. Quorum supports an eth\_sendTransactionAsync API, but does not support the ``restriction" parameter. Given the ``restriction" parameter is an optional parameter, this requirement EE-5.3.1b-JsonRpcTransactionAsyncExt is almost complied with, but not quite. EE-5.3.1c-JsonRpcUnimplemented is also not complied with as it relates to actions taken in regard to the ``restriction" parameter.

EE-5.3.2a-InterChainInteraction: Quorum allows private smart contracts to call functions on public smart contracts which read from public state. This is not an inter-chain interaction as there is only one blockchain upon which both private and public smart contracts sit. As such, Quorum does not offer any inter-chain mediation itself. However, so called, ``Layer 2" solutions, which use the Quorum platform, but do not alter the protocol, could be used to offer some cross-chain support. For example, a smart contract could be written to act as an intermediary between another blockchain and an instance of Quorum. 

EE-5.3.3a-Oracles: Quorum could be configured to interact with a real world oracle. For example, the Oracle could be stored in a secure enclave, and sign transactions from inside the secure enclave \cite{zhang2016}. An Oracle could read information from the blockchain.

\subsubsection{Privacy and Scaling Layer}: The paragraphs below analyse the privacy and scaling related requirements.

EE-6.1.1a-OnChainSecurity: Quorum is a fork of geth. As such, as improved on-chain security techniques become available in geth, they will become available in Quorum, when the Quorum maintainers upgrade Quorum to the version of geth which has the desired feature. For example, Quorum users gained access to zkSnark support when Quorum was updated to include the geth Byzantium release which implements the ECC scalar multiplication, ECC point addition, and pairing checks,  precompiled contracts \cite{byzantium2017}. 

Quorum implements restricted private transactions where the payload data is transmitted to and readable only by the direct participants of a transaction, but not unrestricted private transactions. As such, requirement EE-6.1.2a-PrivTransMethods is complied with, and requirements EE-6.1.2i-UnrestrictedRecipientMasking to EE-6.1.2q-UnrestrictedTransactions are not applicable.

\label{EE-6.1.2b-RestrictedPayloadMaskingStored}EE-6.1.2b-RestrictedPayloadMaskingStored: Constellation stores private transactions in encrypted form. To execute a private transaction, the private transaction payload is passed to Quorum's Golang code, which executes the transaction in the Ethereum Virtual Machine (EVM). The updated state as a result of the transaction is stored in plaintext in the Quorum node's state database. As such, Quorum complies with the requirement to``support masking or obfuscation of the payload data when stored in restricted private transactions (for example, using cryptographic encryption)". It could be argued that this does not meet then intention of the requirement and thus would not match customer expectations. Requirement EE11-1c-PrivateStateAuthenticatedEncryption (section \ref{BC-1d-PrivateStateAuthenticatedEncryption}) has been added to specify that private state that is persisted must be protected.

EE-6.1.2c-RestrictedPayloadMaskingTransit: Constellation encrypts data when it is in transit. As such, Quorum complies with this requirement.

Requirements EE-6.1.2d-RestrictedMetadataMaskingStored and EE-6.1.2e-RestrictedMetadataMaskingTransit relate to transaction metadata. Transaction metadata is defined in the Enterprise Ethereum Client Specification as, ``The set of data that describes and gives information about the payload data in a transaction." Transaction metadata which is stored on the Quorum public state for private transaction are the account which submitted the transaction (the ``from" transaction field), which contract is the target of the transaction (the ``to" field), and the number of transactions the account has previously submitted (the ``nonce" transaction field). Quorum does not have capabilities to protect this metadata which could relate to a private smart contract between one set of participants from other participants who are on the Quorum network. As such, Quorum complies with neither of these requirements.

Quorum does not support relay nodes. As such requirements EE-6.1.2f-RestrictedPayloadRelayStore and EE-6.1.2g-RestrictedMetadataRelayStore are not applicable. Quorum does not support the JSON RPC API eth\_sendTransactionAsync call, and as such requirement EE-6.1.2h-RestrictedDefaultSecure is not applicable.

EE-6.1.2r-PrivateTransactionAddParticipants: The Enterprise Ethereum Client Specification \cite{enteth10} says that, ``Implementations SHOULD be able to extend the set of participants in a private transaction (or forward the private transaction in some way)." Private contracts in Quorum are established by a participant submitting a private transaction using an Ethereum account on a Quorum node, and having that private transaction configured to be ``privateFor" the nodes which belong to the other parties which are to become participants in the contract. Once the contract is deployed, there is no facility to add participants to the contract. Private transactions pertain to executing transactions on a private smart contract. As such, there is no mechanism to extend the set of participants for a private transaction in Quorum.

EE-6.1.2s-PrivateTransactionConsensus: Quorum's public blockchain uses consensus algorithms to achieve consensus. There is however no consensus algorithm used for participants of private transactions to achieve consensus. Private transactions are sent individually to participants of a private contract using Constellation. As such, Quorum does not comply with this requirement. Having a shared state between participants of a private contract is fundamental to blockchain systems. As such, this requirement, though specified as SHOULD, should be marked as MUST.

EE-6.1.3a-OffchainTrustedExecution: Quorum does not support off-chain, trusted execution of transactions and smart contracts. As such, Quorum does not comply with this requirement.

EE-6.2.1a-ImprovedOnchainProcessing: Ethereum Layer 2 scaling techniques are techniques which improve the processing rate of Ethereum without requiring changes to Ethereum's underlying protocols. For example, conceptually, Plasma \cite{poon2017} provides the possibility of executing many parallel blockchains. The Ethereum Layer 2 solutions which become available would allow the Quorum public chain to scale, which would mean that more transactions include private transaction payload hashes could be processed. This would allow a large Quorum network to scale, assuming the participants involved in the private transactions were not overlapping. That is, Quorum could scale by having multiple parallel smart contracts with non-overlapping participants. However, these techniques would not facilitate the transaction rates for a particular contract to scale. This requirement is complied with because there is a way that Ethereum Layer 2 scaling techniques can be used to scale Quorum's transaction processing capabilities. 

EE-6.2.2a-OffchainProcessing: Quorum does not provide the ability for off-chain processing of transactions. There is one Quorum public transaction with the message digest of the private transaction's payload for each private transaction. Additionally, the smart contract function call which is executed as a result of the private transaction is executed in the same Quorum process as the public transactions. As such, Quorum does not comply with this requirement.

EE-6.2.3a-ArchivePrivateState: Quorum does not support any private state data archiving and as such does not comply with this requirement.

EE-6.2.3d-NewGenesisBlock: The genesis block can only be specified once in Quorum. As such, it is not possible for network operators to designate a new genesis block to keep the blockchain size from growing perpetually.

\subsubsection{Core Blockchain Layer} The paragraphs below analyse the core blockchain requirements.

EE-7.1a-StoragePubEth: Quorum is a fork of geth. As such, Quorum has the same storage capability as geth, a public Ethereum client. Quorum complies with this requirement.

EE-7.1b-StorageForOptionalOffchain: Quorum does not implement off-chain operations and as such does not have data storage for such operations.

EE-7.1c-SeparateStoragePerNetwork: Quorum supports a single consortium network. The Quorum public chain and private transactions are in fact the one network. As such, this requirement does not apply. If, however, it was argued that the Quorum public blockchain and private transactions were separate networks for the purposes of this question, then Quorum would not comply with this requirement as Quorum stores public and private state in the same database.

EE-7.1d-DataAccessSameParticipants: Quorum deploys private contracts by sending a private transaction to create a contract to the nodes which should participate in the contract. Participants are free to deploy private contracts which call functions in any other private contract, irrespective of whether the list of participants for the contracts is the same or different. If the participant list of the called contract is smaller than the calling contract, then when a transaction is processed which contains a call from the calling contract to the called contract, the transaction succeeds on nodes which have both contracts and fails on nodes which only have the calling contract. Given this analysis, Quorum complies with requirement EE-7.1d-DataAccessSameParticipants, that a ``...smart contract operating on private state SHOULD be permitted to access private state created by other smart contracts involving the same participants", but does not comply with requirement EE-7.1e-DataAccessDifferentParticipants, that a ``...smart contract operating on private state MUST NOT be permitted to access private state created by other smart contracts involving different participants".

\label{quorum-EE-7.1f-FileDecentralizedStorage}EE-7.1f-FileDecentralizedStorage: IPFS \cite{ipfs} and Swarm \cite{Tron2016} \cite{Tron2017} are the major decentralized storage techniques discussed in the context of Ethereum. Both IPFS and Swarm operate off-chain. Both technologies are at Proof of Concept phase of development, with IPFS being more mature. The requirement words, ``no artificial off-chain file-storage add-ons are needed", rules out both. As such, Quorum does not comply with this requirement.

Quorum supports the standard EVM opcodes and does not have any additional EVM opcodes. As such, requirement EE-7.2a-EvmOpCodes is complied with and requirement EE-7.2b-EvmExtendedOpCodes is not.

EE-7.2c-PublicStateSync: This requirement relates to a node's ability to synchronise public state with public Ethereum nodes on Ethereum MainNet. Quorum is not able to synchronise with Ethereum MainNet as the transaction format used by Quorum is not compatible with Ethereum MainNet. Quorum adds a bit flag to the ``V" field of transactions to indicate if a transaction is private or public. Adding this bit flag to the ``V" field means that Ethereum MainNet clients will not be able to process transactions form Quorum nodes. Quorum is not compliant with this requirement.

EE-7.2d-PrecompiledContracts: This requirement requires that implementations allow users to submit contracts to be compiled and stored as precompiled contracts which can be executed later. Quorum does not provide this capability.

Requirements EE-7.2e-TEE and EE-7.2f-TEEConfigurableEncryption relate to executing contracts in Trusted Execution Environments (TEE). Quorum does not support the use of TEEs and hence does not comply with these requirements.

Requirements EE-7.2.1a-Finality to EE-7.3k-ConsensusConfig relate to consensus algorithms. Quorum currently supports three Proof of Authority (PoA) consensus algorithms: Quorum Chain Consensus  \cite{quorum-wiki} (now deprecated), Raft \cite{raft-paper}, and Istanbul Byzantine Fault Tolerance (IBFT) \cite{ibft}. The consensus algorithm is configurable at start-up. Once the algorithm has been set, it can't be changed. For each of the PoA algorithms, transactions are considered final once they are included in a block. Quorum nodes can not form consensus on Ethereum MainNet as Quorum does not support the Proof of Work consensus algorithm used on Ethereum MainNet. As such, Quorum complies with requirements EE-7.2.1a-Finality, EE-7.3b-MultipleConsensusAlgorithms, EE-7.3c-PrivateConsensusAlg, EE-7.3f-ConsensusAlgDocumented, EE-7.3g-ConsensusAlgModularCon, EE-7.3i-ConsensusIBFT, EE-7.3j-ConsensusOther and does not comply with requirements EE-7.3a-MainNetConsensus and  EE-7.3d-MainNetConsensusAlg.

EE-7.3e-SidechainConsensusAlg: Any PoA, such as IBFT consensus algorithm should allow sidechain clients to form consensus on a sidechain. Quorum complies with this requirement.

EE-7.3h-ConsensusInOutOfBand: Quorum supports in-band communications for consensus, and as such supports this requirement.

EE-7.3k-ConsensusConfig: This requirement indicates that the consensus algorithm for public, private, and sidechains should be configurable. Given Quorum only supports private networks, and is able to configure the consensus algorithm for that network, it is deemed to support this requirement.

\subsubsection{Network Layer} Quorum uses the standard Ethereum DEV Peer to Peer (DEVp2p) protocol \cite{devp2p} for communications between nodes for the ``public" chain. Over the DEVp2p wire protocol, Quorum modifies the Eth protocol to have the ``private" flag in the ``V" field, and other changes to support IBFT. Quorum uses a custom, undocumented, communications protocol to communicate between Constellation nodes, which here will be named, ``Constellation Protocol". Constellation Protocol does not use DEVp2p as the wire protocol.

EE-8.1a-Enode: Nodes are identified and advertised using the Ethereum enode URL format [enode]. Quorum is compliant with this requirement.

The intent of EE-8.1b-DevP2P is that all inter-node communications should use the DEVp2p protocol as the wire protocol which higher layer protocols use. Constellation Protocol does not use DEVp2p as its wire protocol, and as such Quorum is not compliant with this requirement.

EE-8.1c-Eth62Eth63: Quorum does not support eth/62 and eth/63 versions of the Ethereum protocol for the ``public" blockchain. It supports protocols which are similar, but not the same. As such, Quorum does not comply with this requirement.

Requirement EE-8.1d-NewProtocols indicates that implementations may add new high level Ethereum protocols to work on top of the DEVp2p protocol. Quorum's modifications to the Eth protocol could be deemed to be such extensions. As such, Quorum does not comply with this requirement.

Quorum does not support relaying of private data via proxy nodes. As such, it does not comply with requirement EE-8.1e-RelayNodes.

\subsubsection{Anti-Spam} Quorum does not provide any anti-spam mechanisms so attacking nodes can be stopped. As such, Quorum does not comply with this requirement.

\subsubsection{Cross-client Compatibility} The requirements in this section relates to compatibility with Ethereum MainNet and standard APIs. 

EE-10a-PublicEthCompatibility states that, ``Enterprise Ethereum clients SHOULD be compatible with the public Ethereum network to the greatest extent possible." The words, ``to the greatest extent possible" are ambiguous. The intent of the requirement was that Enterprise Ethereum clients should be public chain compatible, and offer a superset of features. Quorum is not MainNet compatible, and as such, does not comply with this requirement.

EE-10b-ExtendedApisSuperset: Quorum adds APIs to submit private transactions, over and above the public Ethereum APIs. These APIs are added such that they are a superset of the public Ethereum APIs. As such, Quorum complies with this requirement.

\subsubsection{Synchronisation and Disaster Recovery}

EE-11a-FastSync: As Quorum is a fork of geth, it supports the same fast synchronisation mode that standard geth does. This can be used to synchronise the public state. As the new client can not be a party to existing contracts, there will be no private state to be synchronised. As such, Quorum supports this requirement.

EE-11b-BackupRestore: Quorum does not support a mechanism for backing up and restoring nodes. A Quorum node could be taken off-line, backed up, and then restarted, however that is not the intention of this requirement. The intention of this requirement is for nodes to be able to be backed-up whilst they are operational. This requirement is particularly important to Quorum as each node in a Quorum network is likely to hold different private state, as each node is likely to have been involved in different private transactions. As such, each Quorum node must back-up its private state or its map of message digests to private transaction encrypted payload. An alternative is that a Quorum node which has experienced a failure, when it comes back on-line, can request that all nodes which previously communicated with it re-submit all previous transactions. There is however no guarantee that all nodes will respond with all previous transactions.

\subsubsection{Additional Blockchain Requirements} The following paragraphs analyse how Quorum meets the additional blockchain requirements.

Quorum does not implement any API permissioning. As such, it does not comply with requirements BC-1a-ApiCallPermissioning, BC-1b-EthereumAccountWhitelist and BC1c-TransactionTypePermissioning.

Quorum does not encrypt the private state which is the result of transactions being executed. As such, Quorum does not comply with requirement BC-1d-PrivateStateAuthenticatedEncryption.

BC-2a-OrganisationallyAwareConsensus: No existing consensus algorithms are explicitly organisationally aware. IBFT and other algorithms could possibly be operated such that each organisation operates one validator node, and have any other nodes they operate as observer nodes. If an organisation's validator node failed, they organisation could detect this and then nominate another node as their observer node. This however, does not provide the automatic redundancy typically associated with blockchain systems. Given this, Quorum is not deemed to support this requirement.

BC-3a-DiscoverableBootstrapInfo: Quorum does not provide any mechanism to discover new parties. All configuration information must be shared out of band prior to a participant becoming a party to a consortium network.

BC-4a-ArchitecturalDecentralization: The private encrypted transaction payloads for a node only exist on that node. This information must be backed-up. If a node fails, other nodes can be asked to resend private transaction payloads. There is however, no guarantee that nodes will resend those transactions. As such, Quorum is architecturally centralised, and does not meet this requirement.

BC-4b-PoliticalDecentralization: Quorum does not have any political centralisation points. Beyond the nodes themselves, which are operated by individual organisations, there are no political centralisation points.

Quorum supports Whisper. This allows node to one or more node and broadcast communications. Whisper includes an anti-spam mechanism such that prior to a message being forwarded by a node, the node must complete a mini-Proof of Work style time hard problem. As such, Quorum complies with requirements BC-5a-OffchainOrgToOrg, BC-5b-OffchainAll, BC-5c-OffchianAntiSpam, and BC-5d-Whisper.

\subsubsection{Additional Sidechain Requirements} Quorum does not support sidechains and as such does not implement the sidechain requirements. Quorum does pin private transactions, and hence the private state, to Quorum's public chain when it sends a transaction onto the public network. However, this is how Quorum pins its private smart contract state to its blockchain, and is not related to sidechain pinning.

If Quorum is used in conjunction with Kaleido's Relay technology \cite{kaleido-relay}, then the state of the Quorum blockchain can be pinned to Ethereum MainNet. Pinning participants can be shielded and the pinning rate can be configured. A method of contesting pins is proposed. The transaction rate is however not shielded and there is no concept of a cipher text observer. As such requirements SC-4a-Pinning, SC-4b-PinningParticipantShielding, SC-4d-PinningContesting, and SC-4f-PinningConfiguration are complied with assuming the use of Kaleido's Relay technology.

\subsection{Parity Technologies' Private Transactions}
\subsubsection{Introduction}
Parity Technologies has created a private transaction and smart contract capability in which a private contract's code and state are stored encrypted inside a public contract. The public contract could be stored on a consortium chain or on Ethereum MainNet. The private transaction capabilities sits on top of Parity's existing Parity Ethereum Client.

To perform a private transaction, a sender creates a transaction with the required function call on the private contract. They send that transaction to a list of validators. The validators each get the encrypted state and code of the private contract from the public contract. They decrypt the code and state, and execute the function which was in the transaction submitted by the sender. They then encrypt the updated state and sign the updated state and return this to the sender. The sender then submits a transaction on the public contract to update the encrypted state of the private contract. The private state is updated if all of the validators have signed the updated encrypted state.

The following sections analyse how Parity Technologies' Private Transactions meets the Enterprise Ethereum Client Specification requirements and the additional Sidechain requirements. The analysis undertaken has been based on Parity's 1.11Beta documentation for Private Transactions \cite{parity-private-transactions}, Parity's Permissioning Model documentation \cite{parity-permissioning}, Parity's private transaction contract source code and test code \cite{parity-github}, and Parity's main codebase on GitHub \cite{parity-github-main}.

\subsubsection{Application Layer}
As Parity is an Ethereum client which supports standard Ethereum APIs and protocols, it allows users to make use of many of the tools which have been and continue to be created for use with Ethereum Clients. The paragraphs below analyse the Application Layer requirements.

EE-4.1a-DApp: Parity does not support the JSON-RPC API Enterprise Ethereum extension described in EE-5.3.1b-JsonRpcTransactionAsyncExt, and as such does not comply with this requirement.

EE-4.3a-Tools: Parity's private contracts could be debugged by initially developing them as public contracts. No specific deployment and debugging tools for Parity's Private Contracts are described in the documentation. In particular, no tools appear to exist for creating private transactions or deploying private contracts. As such, Parity is not deemed to comply with the requirement to ``provide deployment and debugging tools for Enterprise Ethereum smart contracts".

EE-4.3b- FormalVerification: Any formal verification methods which become available for use with standard Ethereum are likely to work with Enterprise Ethereum contracts in Parity. As such, Parity should be deemed to comply with this requirement.

\subsubsection{Tooling Layer} Parity's Private Transaction capability does not use a network of nodes which communicate using peer-to-peer communications. It uses a static list of validator nodes which are specified programmatically at start-up. In the current implementation, this list of nodes can not be updated. Given this, Parity complies with requirement EE-5.1.1a-StaticStartUp which requires a static list of start-up nodes, requirement EE-5.1.1c-WhitelistNodes which requires implementations have the ability to list permissioned nodes, and EE-5.1.1e-WhitelistViaAPI which requires the list of nodes to be able to be specified programmatically. 

As Parity does not run a peer-to-peer network of Validator nodes, the concept of discovery and blacklisting nodes does not make sense. As such, requirements EE-5.1.1b-DisableDiscovery, EE-5.1.1d-BlacklistNodes, and EE-5.1.1f-BlacklistViaAPI are not applicable. 

EE-5.1.1g-CertifyNodes: There is no way to certify the identity of nodes. As such, Parity does not comply with this requirement.

EE-5.1.1h-Organization: Parity does not provide any mechanism to group validator nodes according to which organisation they belong to. As such, Parity does not comply with this requirement.

Parity does not provide any mechanism to authenticate users of the JSON RPC API. The technology stack shown on Parity's website indicates that they plan to create a Secure Data Access Control System and an ID Verification system. However, there is no documentation or examples to explain these technologies. The inference that the author has made is that these technologies are planned, but do not currently exist. As such, there is no ability to whitelist or blacklist participant identities who are permitted to submit transactions or use any other API. Parity does not comply with requirements EE-5.1.2a-WhitelistParticipants, EE-5.1.2b-BlacklistParticipants, EE-5.1.2c-WhitelistParticipantsViaAPI, EE-5.1.2e-CertifyParticipants, and EE-5.1.2f-GroupsRoles. The requirement EE-5.1.2d-BlacklistParticipantsViaAPI is conditional on the blacklisting support from requirement EE-5.1.2b-BlacklistParticipants. As such, this requirement is not applicable. Ethereum Accounts submitting transactions can be permissioned based on transaction type \cite{parity-permissioning}. This type of permissioning is covered by the requirement BC1c-TransactionTypePermissioning; see section \ref{BC1c-TransactionTypePermissioning}.

EE-5.1.3a-SmartContractPermissioning: Parity allows contracts written in any Ethereum Virtual Machine language. This allows application developers to write their contracts such that they use address based permissioning in their smart contracts. This allows Parity to comply with this requirement.

EE-5.1.3b-RuntimeConfigUpdate: Parity's private transaction wrapping contract, as written, does not offer the ability to update the list of validators \cite{parity-private-contract-github}. As such, as written, Parity's Private Transaction code does not comply with this requirement.

EE-5.1.3c-ConfigOptions: Parity's permissioning is set-up using contracts, with the contracts being referenced from a configuration file. The private contract capability is all configured at start-up time via a contract call. As such, Parity's Private Transaction capability does not, "provide configuration through the use of flat files, command-line options, or configuration management system interfaces."

EE-5.1.3d-LocalKeyManagement: Parity offers its Parity Secret Store technology to store keys locally using a threshold encryption scheme \cite{parity-secret-store}. It does not appear to support external key generation and storage or Hardware Security Modules. As such, Parity complies with EE-5.1.3d-LocalKeyManagement but does not comply with EE-5.1.3e-SecureExternalKeyGenStore and EE-5.1.3f-HardwareSecurityModules.

The paragraphs below analyse the integration and deployment tools related requirements.

EE-5.2.1a-IntegrationLibraries: Web3j \cite{web3j-github} has support for some Parity specific APIs, but not for Parity's Private Transaction specific functions. However, Parity's own library \cite{parity-library} does have support for these functions. Parity complies with this requirement.

Parity does not support enterprise software deployment and configuration systems, software fault reporting capabilities, performance management capabilities, security management interaction capabilities, historical analysis, or enterprise management systems. As such, the following requirements are not supported: EE-5.2.2a-EntDeployment, EE-5.2.2b-EntFaultReporting, EE-5.2.2c-EntPerformanceManage, EE-5.2.2d-EntSecurity, EE-5.2.2e-EntHistoricalAnalysis, EE-5.2.2f-EntManagementSystems.

Parity supports the public Ethereum JSON-RPC API but does not support the eth\_sendTransactionAsync API. As such, it complies with requirement EE-5.3.1a-JsonRpcPublicEth but does not comply with EE-5.3.1b-JsonRpcTransactionAsyncExt or  EE-5.3.1c-JsonRpcUnimplemented.

EE-5.3.2a-InterChainInteraction: Parity supports a technology called Parity Bridge \cite{parity-bridge} which allows for the creation of ERC20 tokens on a PoA chain which are backed by Ether on Ethereum MainNet. This inter-chain technology does not provide ``inter-chain mediation capabilities" for Parity's Private Transaction capabilities however. As such, Parity does not comply with this requirement.

EE-5.3.3a-Oracles: An Oracle could execute inside a secure enclave. It could act as a private transaction submitter, and call private contract APIs. As such, Parity complies with this requirement.

The paragraphs below analyse the privacy and scaling related requirements.

EE-6.1.1a-OnChainSecurity: Parity is able to support improved on-chain security as it becomes available. For example, Parity users gained access to zkSnark support when the Byzentium release which implements the ECC scalar multiplication, ECC point addition, and pairing checks, precompiled contracts \cite{byzantium2017} was released.

EE-6.1.2a-PrivTransMethods: Parity private transaction senders communicate private transaction payloads directly to individual validators. The validators update the state of the private contract and return the signed encrypted updated state. The transaction sender submits a public transaction with the updated encrypted state and signatures of all validators to update the private contract's state. If the validators are viewed as participants of the contract, in addition to the transaction submitters, then the  private transaction method can be viewed as restricted private transactions. Though the encrypted state is communicated to the entire blockchain, the private transaction payload is only transmitted to and readable by direct participants of the transaction. Given Parity implements private restricted transactions, requirements EE-6.1.2i-UnrestrictedRecipientMasking to EE-6.1.2q-UnrestrictedTransactions are Not Applicable. Given Parity does not store private transaction payloads, requirements EE-6.1.2b-RestrictedPayloadMaskingStored and EE-6.1.2d-RestrictedMetadataMaskingStored related to private transaction payload storage are not applicable. Requirements EE-6.1.2f-RestrictedPayloadRelayStore and EE-6.1.2g-RestrictedMetadataTransitStore related to relay nodes are also not applicable as Parity's Private Transactions do not use relay nodes.

Private Transactions are communicated directly with  Validator nodes over TLS. As such, Parity complies with the requirements EE-6.1.2c-RestrictedPayloadMaskingTransit and EE-6.1.2e-RestrictedMetadataMaskingTransit to encrypt the payload and metadata whilst in transit.
 
Requirement EE-6.1.2h-RestrictedDefaultSecure is not applicable as the JSON RPC API eth\_sendTransactionAsync call is not implemented.

EE-6.1.2r-PrivateTransactionAddParticipants: The list of nodes which can update the Parity Private Contract is the list of Validators. There is no way to add a new validator to a private contract. As such, Parity does not comply with this requirement.

EE-6.1.2s-PrivateTransactionConsensus: All validators must agree on any update to the private contract state, thus achieving consensus on all private transactions. Parity complies with this requirement.

EE-6.1.3a-OffchainTrustedExecution: Parity does not support off-chain execution in a trusted execution environment. As such, Parity does not comply with this requirement.

\label{Parity-EE-6.2.1a-ImprovedOnchainProcessing}
EE-6.2.1a-ImprovedOnchainProcessing: When improved on-chain processing becomes available, more transactions to update the private state will be able to be executed. As each contract can only have one state change per block, this would have to take the form of more private contracts executing, rather than the one contract executing more transactions. As such, this requirement is deemed complied with, though with the caveat that the limitation of one transaction per block means that to take advantage of the improved on-chain processing users need to have more private smart contracts. 

EE-6.2.2a-OffchainProcessing: Each execution of a private transaction in Parity is done off-chain. The encrypted state is read from the distributed ledger using a View call and the encrypted updated state is written back using a standard transaction. The processing of the private transactions themselves are done in Validators nodes. These Validator nodes do not need to be nodes which are processing transactions on the blockchain. As such, this requirement is complied with because the private transaction process is done off-chain.

EE-6.2.3a-ArchivePrivateState: Parity does not explicitly provide for the ability to archive private state. That said, it could be imagined that the encrypted state could be read, archived, and replaced with a ``dummy" empty updated encrypted state. Hence, Parity has the ability to implement this feature.

EE-6.2.3d-NewGenesisBlock: There is no genesis block used with Parity Private Transactions technology. As such, this requirement is not applicable.

\subsubsection{Core Blockchain Layer} The paragraphs below analyse the core blockchain requirements.

EE-7.1a-StoragePubEth: Parity is a public Ethereum client. The data storage available to the public Ethereum client are available to the private transaction capability. As such, Parity complies with this requirement.

EE-7.1b-StorageForOptionalOffchain: Parity does not implement off-chain storage for the private transactions.

EE-7.1c-SeparateStoragePerNetwork: Parity stores the private contract state on the public chain. As such, it does not comply with the requirement to store private ``network" data separately from other private ``network" data and separately from public network data.

Parity Private Contracts can not access the private state of other contracts. As such, Parity does not comply with requirement EE-7.1d-DataAccessSameParticipants to be able to access private data for private contracts with the same network participants, but does comply with the requirement EE-7.1e-DataAccessDifferentParticipants to not allow access in cases of different participants.

EE-7.1f-FileDecentralizedStorage: As per the analysis in section \ref{quorum-EE-7.1f-FileDecentralizedStorage} for Quorum, there are no existing on-chain decentralised file storage technologies. As such, Parity does not comply with this requirement.

Parity supports the standard EVM opcodes and does not have any additional EVM opcodes. As such, requirement EE-7.2a-EvmOpCodes is complied with and requirement EE-7.2b-EvmExtendedOpCodes is not.

EE-7.2c-PublicStateSync: Parity supports the ability to synchronise the public state with other public Ethereum nodes.

EE-7.2d-PrecompiledContracts: This requirement requires that implementations allow users to submit contracts to be compiled and stored as precompiled contracts which can be executed later. Parity does not provide this capability.

Requirements EE-7.2e-TEE and EE-7.2f-TEEConfigurableEncryption relate to executing contracts in Trusted Execution Environments (TEE). Parity does not support the use of TEEs and hence does not comply with these requirements.

\subsubsection{Execution Sublayer}
Requirements EE-7.2.1a-Finality to EE-7.3k-ConsensusConfig relate to consensus algorithms. Parity requires all Validator nodes to agree on a private state change as a result of a private transaction, and then the resulting private state update is included in the public blockchain based on the consensus algorithm being used on the public blockchain. The public blockchain supports Ethash Proof of Work (PoW), the algorithm used to establish consensus on Ethereum MainNet, and two PoA algorithms: Authority Round (Aura) and Tendermint (experimental only) \cite{parity-consensus}. As such, Parity complies with the requirements EE-7.3a-MainNetConsensus and EE-7.3d-MainNetConsensusAlg related to forming consensus on Ethereum MainNet.

EE-7.2.1a-Finality: When Aura is used, the finality can be determined based on the step duration, number of validator nodes, and the number of faulty nodes allowed. As such, this requirement is complied with.

The requirement EE-7.3b-MultipleConsensusAlgorithms, appears to be complied with, given Ethash, Aura, and Tendermint are supported. However, there is only one algorithm supported for private transaction consensus. As this algorithm has to be used in conjunction with the other algorithms, Parity does not comply with this requirement.

EE-7.3c-PrivateConsensusAlg: This requirement is complied with as algorithms which allow consensus to be formed on a consortium network, namely Aura, are supported. The Aura consensus algorithm could be used on a sidechain, and hence requirement EE-7.3e-SidechainConsensusAlg is supported.

EE-7.3f-ConsensusAlgDocumented: The Ethash, Aura, and Tendermint algorithms are documented \cite{parity-consensus}. As such, Parity conforms with this requirement.

EE-7.3g-ConsensusAlgModularConf: The implementations of each consensus algorithm are in separate directories of the Parity source tree. Each public algorithm is configurable, with parameters supplied by values in the ``chain config file" which is passed into the \texttt{parity} executable as a command line argument. However, there is no way to configure the Private Transaction consensus algorithm. As such, Parity does not comply with the requirements EE-7.3g-ConsensusAlgModularConf and EE-7.3k-ConsensusConfig which require configurable consensus.

EE-7.3h-ConsensusInOutOfBand: Parity's Private Transactions uses both out-of-band communications for consensus, and as such supports this requirement. Transaction senders communicate directly with Validators, create the signed updated state, and submit that as a transaction. The transaction sender co-ordinates the consensus out-of-band relative to the Ethereum node to Ethereum node communications, and then submits the transaction to the public chain.

EE-7.3i-ConsensusIBFT: Parity's Private Transactions feature does not support Istanbul Byzantine Fault Tolerance (IBFT) consensus algorithm [EIP-650], and as such does not comply with this requirement. It does support other consensus algorithms, so it complies with requirement EE-7.3j-ConsensusOther.

\subsubsection{Network Layer} Requirements EE-8.1a-Enode, EE-8.1b-DevP2P, and EE-8.1c-Eth62Eth63 relate to the implementation using standard Ethereum communications mechanisms and addressing to communicate between nodes. Parity complies with all of these requirements as it is a client which is able to operate on Ethereum MainNet. 

Requirement EE-8.1d-NewProtocols indicates that implementations may add new high level Ethereum protocols to work on top of the DEVp2p protocol. Parity has added some protocols to work on top of DEVp2p such as the Parity Light Protocol. As such, Parity complies with this requirement.

EE-8.1e-RelayNodes: Parity's Private Transactions do not support the concept of relay nodes.

\subsubsection{Anti-Spam} To submit a private state update, a transaction must be submitted on the public blockchain. If this blockchain is Ethereum MainNet, then uses need to spend ``Ether" to buy ``gas". As such, the amount of spurious transactions created by attackers or malfunctioning nodes is limited due to the cost of such behaviour. As such, requirement EE-9a-AntiSpam is complied with, assuming an Ethereum MainNet deployment. However, for a consortium deployment, no such economic disincentive exists. As such, Parity partially complies with this requirement.

\subsubsection{Cross-client Compatibility} The requirements in this section relates to compatibility with Ethereum MainNet and standard APIs. Parity is an Ethereum Client capable of operating its Private Transactions on Ethereum MainNet. It extends the features of existing Ethereum APIs. As such, Parity is complies with requirements EE-10a-PublicEthCompatibility and EE-10b-ExtendedApisSuperset.

\subsubsection{Synchronisation and Disaster Recovery} Parity supports the snapshotting of blockchains at specific block numbers. A Parity node can then be started based on that snapshot, and fast synchronised to be up to date using a feature called warp-sync. As such, Parity complies with requirements EE-11a-FastSync and EE-11b-BackupRestore.

\subsubsection{Additional Blockchain Requirements} The following paragraphs analyse how Parity meets the additional blockchain requirements described in this paper.

Parity does not implement any API permissioning at the node level. As such, it does not comply with requirements BC-1a-ApiCallPermissioning and BC-1b-EthereumAccountWhitelist. Parity does allow a transaction type permissioning contract to be deployed which specifies which accounts can execute which type of transaction \cite{parity-permissioning}. As such, Parity complies with requirement BC1c-TransactionTypePermissioning.

Parity encrypts the code and state of a Private Contract using AES/CTR with an AES 128 bit key. AES/CTR is not an authenticated encryption technique. As such, Parity does not comply with requirement BC-1d-PrivateStateAuthenticatedEncryption.

BC-2a-OrganisationallyAwareConsensus: Parity's Private Transaction technology does not support an Organisationally Aware Consensus algorithm, and as such does not comply with this requirement.

BC-3a-DiscoverableBootstrapInfo: Parity does not provide any mechanism to discover new parties. All configuration information must be shared out of band prior to a participant becoming a party to a consortium network.

BC-4a-ArchitecturalDecentralization: Parity's Private Contracts being stored inside public contracts facilitate this solution being architecturally decentralised. The secret stores used to encrypt and decrypt the code and state can be set-up in a decentralised manner. As such, Parity complies with this requirement.

BC-4b-PoliticalDecentralization: Beyond the nodes themselves, which are operated by individual organisations, Parity has no political centralisation points.

Parity's support of the Whisper \cite{parity-whisper} protocol allows node to one or more node and broadcast communications. Whisper includes an anti-spam mechanism such that prior to a message being forwarded by a node, the node must complete a mini-Proof of Work style time hard problem. As such, Parity complies with requirements BC-5a-OffchainOrgToOrg, BC-5b-OffchainAll, BC-5c-OffchianAntiSpam, and BC-5d-Whisper. 

\subsubsection{Additional Sidechain Requirements} Parity does not support sidechains and as such does not implement the sidechain requirements. Parity does store the encrypted state of a private contract onto the public chain. However, this is not the same as cross-chain pinning. In Parity, there is only one blockchain on which both private and public smart contracts reside. There is no pinning of private state from one blockchain onto another chain.

\subsection{Hyperledger Fabric}
\subsubsection{Introduction}
Hyperledger Fabric \cite{hyperledger-fabric} \cite{hyperledger-fabric-doc} \cite{androulaki2018} is a distributed ledger platform originally created by IBM and now hosted by The Linux Foundation. The platform executes smart contracts, called ``chaincode", which can be written in any language, in Docker containers. Private transactions execute within ``channels" such that only the participants of the transaction see the transaction.

Organisations can operate one or more nodes. Each node has an identity in the form of an X.509 certificate. Each organisation operates a Membership Service Provider (MSP) which issues identities for the nodes, TLS root certificates, and certificates for users of the nodes. The MSP uses the Fabric CA to issue certificates and Certificate Revocation Lists (CRL). Anchor Peers are special nodes which can be contacted to find out information about other nodes belonging to an organisation. To establish a network, the IP address of the Anchor Peer and the root CA certificate for the organisation needs to be shared out-of-band with all other organisations which will be on the network. To improve resistance to failure, an organisation could operate more than one Anchor node.

A Hyperledger Fabric channel can be thought of as a ``subnet" within the overall network for performing private transactions between two or more organisations. The Orderer System Channel is a special channel between all Orderer nodes on the network. A channel config transaction which includes information about the Anchor Peers for each organisation involved in a channel is executed on the Orderer System Channel to create an Application Channel. The channel configuration can be updated, for example to add an organisation to the channel, by executing a configuration update transaction.

To process a transaction, a participant submits a proposed transaction to Endorsing nodes. Each endorsing node executes the transaction and returns to the participant signed copies of the state before and after the transaction was executed. Some  endorsing nodes may have out of date state, which means that the returned values from endorsing nodes may be inconsistent. Based on the ``endorsement policy", the participant may need to have all endorsing nodes return the same state, or may be able to proceed with only a subset of endorsements. The participant then submits the signed transaction, with the endorsed before and after states to Ordering nodes. The Ordering nodes come to a consensus on the order of transactions within a block. The block is broadcast to all nodes. All nodes then validate the transactions, checking the endorsement policies and checking that the state before a transaction matches the expected endorsed before state. If the before state is not as expected, then the transaction is marked as invalid. The endorsed after states are applied to the distributed ledger. 

Chaincode may call any other chaincode within a channel. No inter-channel communication is allowed. However, all channels share the same un-encrypted state database, which could allow carefully crafted malicious chaincode from one channel to read state from chaincode from another channel. 

The following sections analyse how Hyperledger Fabric meets the Enterprise Ethereum Client Specification requirements and the additional Sidechain requirements. The analysis undertaken has been based on Hyperledger Fabric's 1.1 documentation \cite{hyperledger-fabric-doc}, Androulaki et al.'s analysis of Hyperledger \cite{androulaki2018}, and discussions with senior technical staff who previously worked on Hyperledger Fabric at IBM. 

\subsubsection{Application Layer} EE-4.3a-Tools: Hyperledger Composer \cite{hyperledger-fabric-composer} is a tool which allows for the creation and deployment of Hyperledger Fabric Chaincode. There are plans to implement formal verification of business logic in Hyperledger Composer \cite{hyperledger-composer-meet}, but these features are yet to be released. As such, Hyperledger Fabric complies with the requirement to have to tools EE-4.3a-Tools but does not comply with the requirement to have formal verification tools, requirement EE-4.3b- FormalVerification.

\subsubsection{Tooling Layer} EE-5.1.1a-StaticStartUp: Hyperledger Fabric requires that a static list of Anchor nodes and other nodes which can be on the network be shared out-of-band. As such, Hyperledger Fabric complies with this requirement to be able to specify a static list of start-up nodes.

EE-5.1.1b-DisableDiscovery: Any node within an organisation can contact one of the organisation's Anchor nodes to discover all nodes which are on the channel. There is no ability to disable this node discovery. As such, Hyperledger Fabric does not comply with this requirement.

In Hyperledger Fabric organisations are whitelisted to be part of a network and to be part of a channel. The organisation can add nodes to be part of the network or channel. This whitelisting is done in part by configuration file and in part via API. Blacklisting is not supported. As such, requirements EE-5.1.1c-WhitelistNodes, and EE-5.1.1e-WhitelistViaAPI are complied with and requirement EE-5.1.1d-BlacklistNodes is not supported which means that requirement EE-5.1.1f-BlacklistViaAPI is not applicable.

EE-5.1.1g-CertifyNodes: Nodes are certified in Hyperledger Fabric using the MSP. As such, it complies with this requirement.

EE-5.1.1h-Organization: Hyperledger Fabric provides the ability to group nodes as being part of an organisation, meeting this requirement.

Participants who interact with a node's API are identified using certificates issued by the MSP. The list participants allowed to interact with an API is limited by a whitelist in the channel configuration. The participants' capabilities can be restricted to being read, write, or admin. The configuration can be specified programmatically.  Blacklisting is not supported. Participants can be assigned roles. As such, requirements EE-5.1.2a-WhitelistParticipants, EE-5.1.2c-WhitelistParticipantsViaAPI, EE-5.1.2e-CertifyParticipants, and EE-5.1.2f-GroupsRoles are complied with. Requirement EE-5.1.2b-BlacklistParticipants is not supported, and requirement EE-5.1.2d-BlacklistParticipantsViaAPI is not applicable.

EE-5.1.3a-SmartContractPermissioning: Chaincode in Hyperledger Fabric can determine which identity created a transaction and as such could implement a ``smart contract-based" permissioning scheme. EE-5.1.3b-RuntimeConfigUpdate: Channel and network configuration can be updated at run time without the need to restart. EE-5.1.3c-ConfigOptions: Hyperledger Fabric can be configured using files and command line options. EE-5.1.3d-LocalKeyManagement: The MSP provides some key management capabilities. The Fabric CA which backs the MSP can be configured to use a Hardware Security Module (HSM). As such, Hyperledger Fabric complies with these requirements.

EE-5.2.1a-IntegrationLibraries: This requirement is complied with as a range of integration libraries exist for Hyperledger Fabric including ones written in the go-lang \cite{hyperledger-fabric-sdk-go} and Java \cite{hyperledger-fabric-sdk-java} programming languages.

Hyperledger Fabric operates nodes and chaincode as Docker containers. Enterprise deployment tools which work with Docker containers could be configured to deploy and monitor deployments of Hyperledger Fabric. Hyperledger Fabric has configurable logging. This logging can be integrated into software fault reporting systems and security management systems. Performance analysis tools such as Performance Traffic Engine \cite{hyperledger-fabric-performance} can be used to measure performance of Hyperledger Fabric in a range of scenarios. Hyperledger Fabric does not support enterprise management systems using Common Management Information Protocol (CMIP), Web-Based Enterprise Management (WBEM), or Application Service Management (ASM) instrumentation. As such, requirements EE-5.2.2a-EntDeployment, EE-5.2.2b-EntFaultReporting, EE-5.2.2c-EntPerformanceManage, EE-5.2.2d-EntSecurity, EE-5.2.2e-EntHistoricalAnalysis are complied with and requirement EE-5.2.2f-EntManagementSystems is not.

EE-5.3.2a-InterChainInteraction: Hyperledger Fabric provides no mechanism for inter-blockchain interaction or for inter-Hyperledger Fabric channel communications. As such, Hyperledger Fabric does not comply with this requirement.

EE-5.3.3a-Oracles: An Oracle could run inside a trusted enclave and then publish values to chaincode in a channel. The Oracle would need to be configured with a valid identity for the channel. In this way the Oracle could securely send and receive data from chaincode.

\subsubsection{Privacy and Scaling Layer}
EE-6.1.1a-OnChainSecurity: As Hyperledger Fabric chaincode is executed in a Docker container, application implementers can choose to use any on-chain security features such as zkSnarks, Range Proofs and other techniques.

EE-6.1.2a-PrivTransMethods: Hyperledger Fabric's channel approach means that private transactions are only transmitted to and readable only by the direct participants of a transaction. As such, the restricted private transaction requirements and not the unrestricted requirements are applicable to Hyperledger Fabric.

Hyperledger Fabric stores transactions and state in plaintext \cite{hyperledger-fabric-doc}. As of version 1.1, chaincode can choose to encrypt data prior to storing state. The keys used for this must be managed outside of Hyperledger Fabric. Hyperledger Fabric communications transactions in plaintext over TLS. Transactions can be encrypted using keys managed outside of Hyperledger Fabric. The metadata however can not be encrypted. As such, Hyperledger Fabric complies with the requirements EE-6.1.2b-RestrictedPayloadMaskingStored and EE-6.1.2c-RestrictedPayloadMaskingTransit, but does not comply with  requirements EE-6.1.2d-RestrictedMetadataMaskingStored and EE-6.1.2e-RestrictedMetadataMaskingTransit.

Hyperledger Fabric does not support relay nodes. As such, requirements EE-6.1.2f-RestrictedPayloadRelayStore and EE-6.1.2g-RestrictedMetadataRelayStore are not applicable.

EE-6.1.2r-PrivateTransactionAddParticipants: Participants can be added to a Hyperledger Fabric channel. This adds the participant to a private transaction, thus complying with this requirement. 

EE-6.1.2s-PrivateTransactionConsensus: Orderers form consensus on the order of transactions in blocks for a channel. As such, Hyperledger Fabric complies with the requirement for nodes involved in a private transaction to be able to form consensus.

EE-6.1.3a-OffchainTrustedExecution: Off-chain trusted execution of Hyperledger Fabric chaincode is not yet supported. It is however an active area of research which may become available in due course \cite{hyperledger-fabric-sgx}.

Hyperledger Fabric executes chaincode in Docker containers. The Docker containers could call out to other containers on other computers to perform computations in parallel. This provides improved processing capabilities which can be viewed as both on-chain and off-chain. As such, Hyperledger Fabric complies with requirements EE-6.2.1a-ImprovedOnchainProcessing and EE-6.2.2a-OffchainProcessing.

EE-6.2.3a-ArchivePrivateState: Hyperledger Fabric allows data to be backed up and restored. As such, Hyperledger Fabric does not comply with this requirement. 

Requirement EE-6.2.3d-NewGenesisBlock relates to limiting the size of the blockchain by specifying a new genesis block. Hyperledger Fabric does not offer this capability.

\subsubsection{Core Blockchain Layer}
EE-7.1b-StorageForOptionalOffchain: Hyperledger Fabric does not provide storage for use with off-chain operations. As such, this requirement is not supported.

EE-7.1c-SeparateStoragePerNetwork: Hyperledger Fabric supports multiple blockchains in the form of channels. All channels are on the same overall consortium network, however transactions on a channel are only between nodes which are a part of the channel. As such, from the perspective of this requirement, Hyperledger Fabric supports multiple networks. Hyperledger Fabric stores the blockchain for each channel in a separate file, however the state for all channels is stored in the same database  \cite{hyperledger-fabric-doc}. As such, Hyperledger Fabric does not comply with this requirement.

EE-7.1d-DataAccessSameParticipants: Hyperledger Fabric allows chaincode operating within the same channel to access other chaincode state in the same channel. As such, Hyperledger Fabric complies with this requirement. 

EE-7.1e-DataAccessDifferentParticipants: There is no facility for chain code in one channel to call chaincode in another channel. All data is stored in a common state database. As such, Hyperledger Fabric does nt comply with this requirement to not ``permit to access private state created by other smart contracts involving different participants". It could conceivably be possible for chaincode on one channel reach the state of chaincode in a different channel.

EE-7.1f-FileDecentralizedStorage: Storing files in Hyperledger Fabric is facilitated by converting files to based64 encoded strings and storing strings. As such, it can store files without, "artificial off-chain file-storage add-ons are needed."

Brandenburger et al. \cite{hyperledger-fabric-sgx} have produced research on operating Hyperledger Fabric chaincode in a Trusted Execution Environments (TEE). This research is however is not something which can be used in a Hyperledger Fabric production environment at the moment. As such, Hyperledger Fabric does not comply with requirements EE-7.2e-TEE or EE-7.2f-TEEConfigurableEncryption.

\label{HyperLedgerFabric-EE-7.2.1a-Finality}
EE-7.2.1a-Finality: Hyperledger Fabric uses Kafka \cite{kafka} as part of the ordering service to provide consensus. Transactions can be considered final once they have been ordered using Kafka, put into a block, and then validated by each node. When a block is being validated by nodes, if a transaction's input state does not match the current input state, then a transaction is rejected and needs to be resubmitted. As such, though this requirement that, ``transactions SHOULD be considered final after a defined interval or event." \cite{enteth10} is correct, participants using Hyperledger Fabric must ensure that transaction included in blocks haven't been rejected, to ensure the transactions have been finalised. 

Hyperledger Fabric is capable of supporting multiple consensus algorithms. However, at present, the only production grade algorithm supported is Kafka. This algorithm could be used for private channels or for sidechains. The consensus algorithm is documented in Androulaki  et al.'s paper \cite{androulaki2018}. Consensus information is communicated in-band. As such, Hyperledger Fabric complies with the requirements EE-7.3b-MultipleConsensusAlgorithms, EE-7.3c-PrivateConsensusAlg, EE-7.3e-SidechainConsensusAlg, EE-7.3f-ConsensusAlgDocumented, EE-7.3h-ConsensusInOutOfBand, and EE-7.3j-ConsensusOther.

EE-7.3g-ConsensusAlgModularConf: The consensus algorithm code is modular in the Hyperledger Fabric codebase. Kafka must be configured with X.509 certificates. As such, Hyperledger Fabric complies with the requirement for the, ``Consensus algorithm implementations SHOULD be modular and configurable".

EE-7.3i-ConsensusIBFT: As Hyperledger Fabric does not support IBFT, this requirement is not met.

EE-7.3k-ConsensusConfig: When a channel is established, the ordering service to be used for the channel is specified. As such, the consensus algorithm is able to be configured for each channel, thus complying with this requirement.

\subsubsection{Network Layer}

EE-8.1e-RelayNodes: Relay nodes are not supported by Hyperledger Fabric.

\subsubsection{Anti-Spam}
EE-9a-AntiSpam: Hyperledger Fabric does not appear to support any anti-spam mechanisms. Hyperledger Fabric assumes all parties on the network are fully trusted. As such, Hyperledger Fabric does not comply with this requirement.

\subsubsection{Synchronisation and Disaster Recovery}
EE-11a-FastSync: Hyperledger Fabric does not appear to support any accelerated synchronisation mechanism to get new nodes synchronised quickly.

EE-11b-BackupRestore: Hyperledger Fabric does not currently support the backing up and restoring of node data. This is however an issue which is actively being discussed \cite{hyperledger-fabric-backup}.

\subsubsection{Additional Enterprise Ethereum Client Requirements} Hyperledger Fabric restricts which accounts can submit which transactions and has API permissioning. For example, to deploy a new chaincode, the chaincode which has been signed by the entities which own the code, must be deployed on a node which is a part of the channel by a local MSP administrator using a signed transaction proposal \cite{hyperledger-fabric-chaincode}. As such, Hyperledger Fabric complies with the requirements BC-1a-ApiCallPermissioning, BC-1b-EthereumAccountWhitelist, and BC1c-TransactionTypePermissioning.

BC-1d-PrivateStateAuthenticatedEncryption: Hyperledger Fabric does not encrypt the blockchain state. As such, it does not comply with this requirement to authenticated-encrypt the blockchain state.

BC-2a-OrganisationallyAwareConsensus: Hyperledger Fabric's Kafka isn't an organisationally aware consensus algorithm. It relies on each organisation only having a single ordering node. As such, Hyperledger Fabric does not comply with this requirement.

BC-3a-DiscoverableBootstrapInfo: Hyperledger Fabric does not offer any ability to discover parties they have no previous relationship with. As such, it does not comply with this requirement.

BC-4a-ArchitecturalDecentralization: Hyperledger Fabric's architecture aims to have no single points of failure. An organisation can have more than one anchor peer, though typically only has one. The anchor peer needs to be contactable for a channel to be created which includes the organisation the anchor belongs to. The gossip protocol is such that each organisation has a leader peer as the contact point for the organisation for the channel. There is however a leader fail-over mechanism. Additionally, each organisation has a certificate authority and Membership Service Provides (MSP) which issue identities. Though Hyperledger Fabric has the ability to have primary and secondary nodes, for example with the anchor peers, this is not the same as decentralisation. If Hyperledger Fabric was a fully decentralised solution, there would be no special peers, or any peer would be able to automatically reconfigure itself to become any peer type which has failed. As such, Hyperledger Fabric does not meet this requirement.

BC-4b-PoliticalDecentralization: Hyperledger Fabric networks and channels are operated by the members of the network. They do not need a central organisation to coordinate activities. As such, Hyperledger Fabric complies with this requirement.

Hyperledger Fabric does not provide an off-chain communications mechanism. As such, it does not comply with requirements BC-5a-OffchainOrgToOrg, BC-5b-OffchainAll, and BC-5c-OffchianAntiSpam.

\subsubsection{Additional Sidechain Requirements} Hyperledger Fabric implements a sidechain feature, which it calls, ``channels".  Identities who are authorised to create channels are set-up programmatically using the MSP. However, there is no blacklisting support. There is also no support to limit which nodes or other organisations can send requests to create a channel. As such, Hyperledger Fabric complies with requirement SC-2a-EstablishmentApiWhitelist, but not SC-1a-EstablishmentNodesWhitelist, SC-1b-EstablishmentNodesBlacklist, and SC-2b-EstablishmentApiBlacklist.

Hyperledger Fabric supports an API to create a channel \cite{hyperledger-fabric-client-code}. However, there does not appear to be an API to find an existing channel. As such the requirement SC-3a-SidechainFindOrEstablishmentApi is partially supported. Hyperledger Fabric APIs take a ``channel" parameter, which is equivalent to the concept of the ``sidechain identifier" parameter. As such, Hyperledger Fabric supports the SC-3b-SidechainIdentifier requirement.

Hyperledger Fabric does not support sidechain state pinning. As such, none of the pinning requirements SC-4a-Pinning to SC-4f-PinningConfiguration are supported.

For sidechain systems to be effective, Ethereum clients which support sidechains need to allow a multitude of sidechains to operate simultaneously. In the example there were three sidechain in addition to the management chain. Ethereum Private Sidechain clients need to have a specific requirement to support multiple clients.

SC-4g-MultipleSidechains: Hyperledger Fabric allows for multiple channels to operate simultaneously. As such, it complies with this requirement.

SC-5a-DataAccessDifferentParticipants: No cross-channel communication is allowed in Hyperledger Fabric. As such, it does not comply with this requirement.

SC-6a-SidechainArchive: Hyperledger Fabric does not provide the facility to archive channels. As such, it does not meet this requirement to be able to archive channels.

\section{Comparison}
Quorum, Parity, and Hyperledger Fabric are very different platforms which approach private smart contracts very differently. Despite those differences, all platforms provide some fundamental features such as only allowing whitelisted nodes to access a private contract (requirement EE-5.1.1c-WhitelistNodes) and allowing for in-contract permissioning (requirement EE-5.1.3a-SmartContractPermissioning). However, there are also many differences, as described in the following paragraphs.

Hyperledger Fabric's assumption that all nodes on the network and all participants are trusted has meant that the only production grade consensus algorithm offered by this platform, Kafka, is not Byzantine Fault Tolerant. Parity's Private Transactions only has a single algorithm which requires all participants to agree, and hence is not Byzantine Fault Tolerant. Quorum offers two production grade consensus algorithms of which one, IBFT is Byzantine Fault Tolerant. 

Protection of state is another point of difference. Parity's Private Transactions' state is stored encrypted. Quorum and Hyperledger Fabric do not encrypt the state. That said, Parity's state is only encrypted with AES/CTR, and not an authenticated-encrypted algorithm such as AES/GCM. As such, none of the platforms meet the requirement BC-1d-PrivateStateAuthenticatedEncryption.

Parity is public chain compatible whereas Quorum is not. Hyperledger Fabric, not being an Ethereum based platform is also not public chain compatible.

Hyperledger Fabric's ``channel" feature is similar to sidechains. This has meant that it supports some of the sidechain requirements. However, importantly, it does not support any inter-channel communications. Neither Quorum nor Parity offer any form of sidechain support.

Hyperledger Fabric's system has the concept of ``organisation", recognising that companies are likely to operate more than one node, and as such meets the requirement EE-5.1.1h-Organization. Neither Quorum nor Parity support the concept of organisation.

Hyperledger Fabric does not support an off-chain secure communications service. However, both Parity and Quorum support the Whisper protocol, and as such support requirements BC-5a-OffchainOrgToOrg, BC-5b-OffchainAll, and BC-5d-Whisper.

None of the platforms support the requirement BC-3a-DiscoverableBootstrapInfo to bootstrap information about an organisation they had not previously interacted with, so that a blockchain can be established on-demand. Additionally none of the platforms support the concept of relay nodes, requirement EE-8.1e-RelayNodes.

All of the platforms were Politically Decentralised, however, only Parity fully fulfilled the requirement to be Architecturally Decentralised.

As can be seen from the above analysis, none of the platforms meet all of the requirements at time of writing. The section below proposes the architecture of a system which meets the requirements laid out in this document.

\section{Solution}
\subsection{Introduction}
This section presents an overview of Ethereum Private Sidechains.

\subsection{Multi-Blockchain Architecture}
\begin{figure}
  \includegraphics[width=\linewidth]{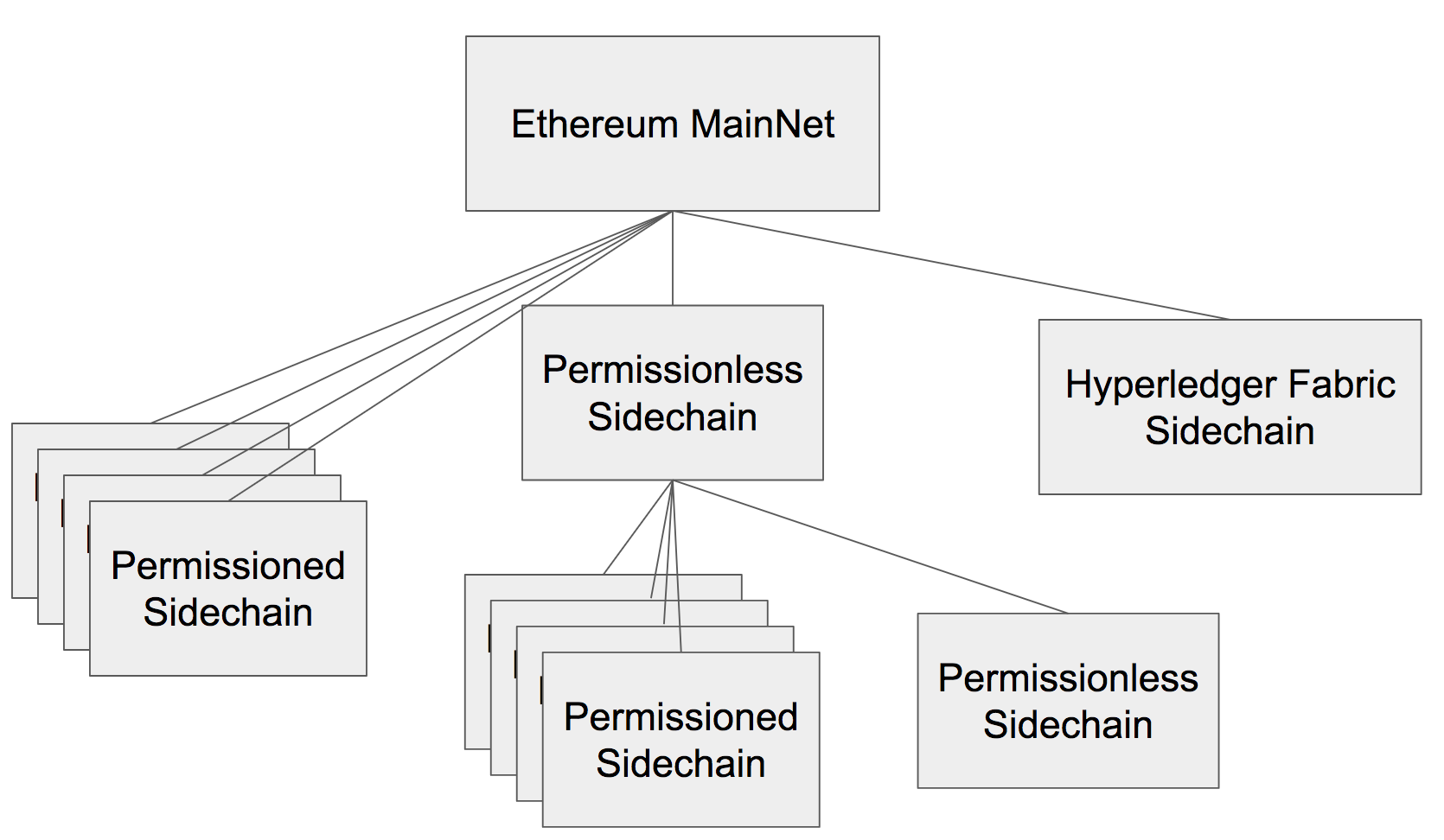}
  \caption{Multi-Blockchain Architecture}
  \label{fig:sidechain-architecture}
\end{figure}

The top level Ethereum Private Sidechain architectural concept is presented in figure \ref{fig:sidechain-architecture}. In this architecture multiple blockchains, known as sidechains, use Ethereum MainNet as a management and coordination blockchain. Though possibly indicating the title Ethereum Private Sidechains is a misnomer, sidechains could be Ethereum based or other blockchain platforms, for instance Hyperledger Fabric channels. Additionally, though the sidechains all have permissioning capabilities, some may be configured to be permissionless, allowing any organisation to connect to the sidechain.

As indicated by the lines in the diagram, some sidechains may use Ethereum MainNet for management and coordination. Other sidechains may wish to have management and coordination activities done on a separate sidechain which is itself in turn managed on Ethereum MainNet. Having this delegation model is advantageous for several reasons: the gas cost on the sidechain could be set such that performing management and coordination activities could be significantly cheaper on the sidechain than on Ethereum MainNet. Additionally, the sidechain could have its permissioning set such that it was permissionless, allowing any organisation to use it, or could be permissioned such that only consortium members could use it. Another benefit of using a delegation model is that it allows transactions which would occur on Ethereum MainNet to be moved to a separate blockchain. This means that the capabilities of Ethereum MainNet can scale and means that the activities of the sidechain can proceed even if MainNet is under heavy load as happened due to the Crypto Kitties game \cite{crypto-kitties} and other games which use Ethereum MainNet \cite{crypto-kitties-other}.

This design allows the solution to meet the requirement to have multiple sidechains (SC-4g-MultipleSidechains). The overall solution would be MainNet compatible as the solution requires a MainNet client to operate as a management and configuration point. This would allow the solution to meet the requirement EE-7.3a-MainNetConsensus. Individual sidechains are unlikely to be MainNet compatible as they are likely to be using Proof of Authority consensus algorithms such as IBFT.

\subsection{Ethereum Registration Authority}
Ethereum Registration Authorities (ERA) are smart contracts which organisations can use to determine bootstrap information about other organisations such that they can establish sidechains with them. The contracts map domain name to the address of a delegate ERA and / or an Organisational Information (OrgInfo) contract. OrgInfo contracts hold a mapping of name-value pairs which can represent arbitrary information, for example the enode addresses of sidechain nodes and encryption keys. Any information currently held on Domain Name servers could be held in these contracts.

Root ERAs are operated by companies which attest to the fact that a domain name matches a company which matches an Ethereum Address. Organisations can host their own Delegate ERAs. An organisation can list the address of their delegate ERA with one or more root ERAs. 

\begin{figure}
  \includegraphics[width=\linewidth]{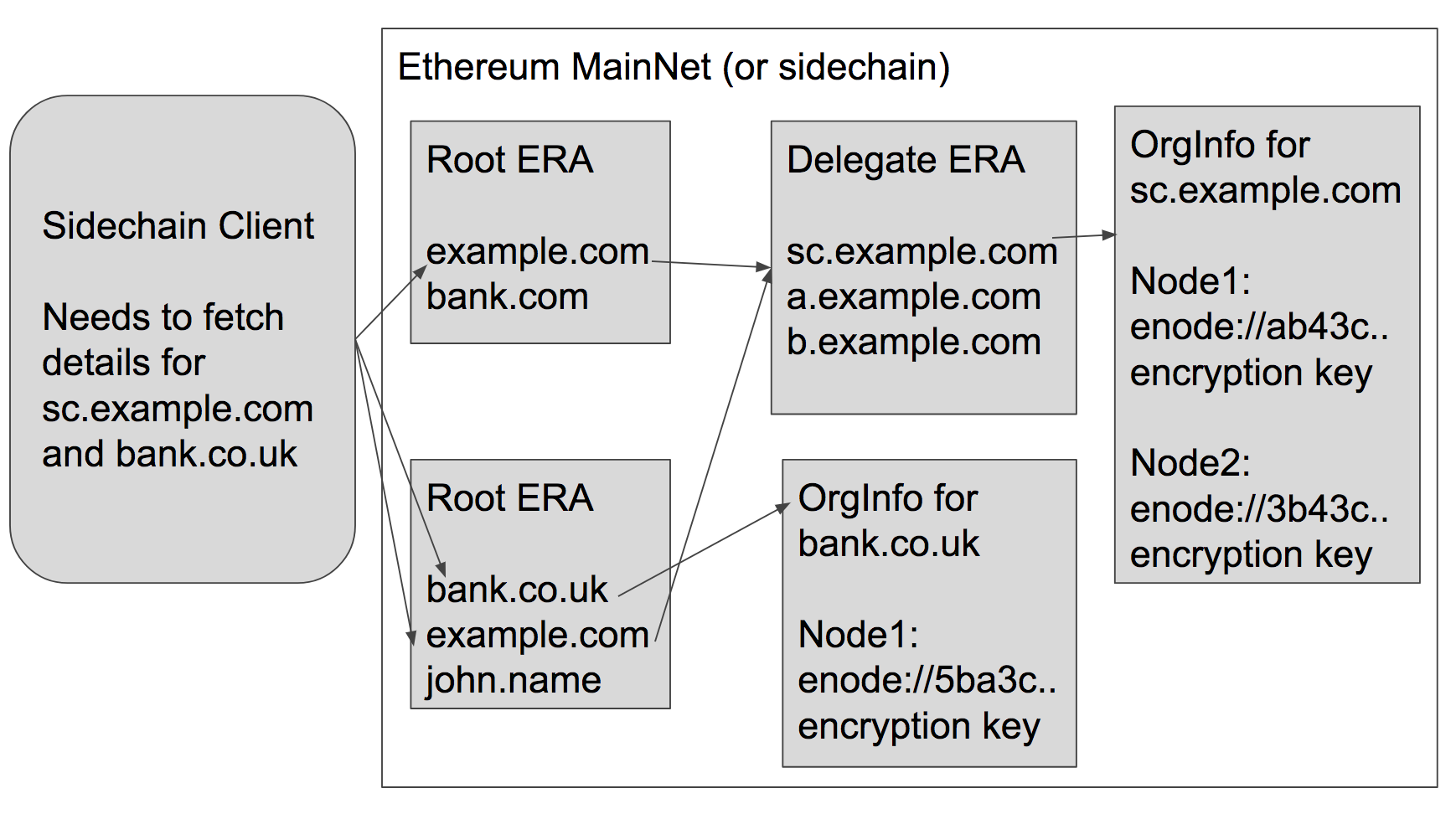}
  \caption{Root and Delegate Ethereum Registration Authorities}
  \label{fig:ethereum-registration-authority}
\end{figure}

In figure \ref{fig:ethereum-registration-authority} there is a sidechain client which wishes to determine bootstrap information for \texttt{sc.example.com} and \texttt{bank.co.uk}. The sidechain client trusts two Root ERAs. \texttt{sc.example.com} is listed with both Root ERAs and operates its own Delegate ERA, whereas  \texttt{bank.co.uk} is only listed in one Root ERA and does not operate its own Delegate ERA. The sidechain client asks each Root ERA for information about the domains. If the Root ERA does not contain information about the domain being searched for, then the sidechain client asks if it has information about the domain's parent or grand parent. That is, if the sidechain client was searching for information about \texttt{aa.bb.sc.example.com}, it would ask the Root ERA if it had information about the domain \texttt{aa.bb.sc.example.com}, the parent domain \texttt{bb.sc.example.com}, the grand parent domain \texttt{aa.bb.sc.example.com}, and so forth. 

Information listed in the OrgInfo contracts is held as a map of names and values. To provide extensibility the names need to be formatted such that they can be uniquely extended. To limit the amount of gas cost, the names should be a consistent single word (256 bits) size. With these ideas in mind, the format for names is a Keccak-256 message digest of the reverse domain order. Using a Keccak-256 message digest means that any name will be exactly one word long. Using reverse domain name order means that an organisation that wishes to create their own names can do so safe in the knowledge that they own the name space within the domain that they control. Hence, \texttt{example.com} could define a name \texttt{com.example.app1.pubkey} which might be for a public key for an application. To provide interoperability, common names should be standardised. For example, perhaps \texttt{org.ethereum.enode} could be for enode addresses. 

Having Root ERAs may appear to mean that this solution will not meet the political decentralisation requirement, BC-4b-PoliticalDecentralization. However, organisations can choose to register their Delegate ERA or OrgInfo contracts in as many Root ERAs as they wish. Additionally, sidechain clients can choose which Root ERAs they trust. ERAs which are known to be reputable are more likely to attract organisations wishing to list with them and are more likely to be used by sidechain clients. This could lead to a degree of political centralisation.

\subsection{Management and Pinning}
To meet the requirement SC-4a-Pinning, this paper proposes a Management and Pinning contract which resides on Ethereum MainNet or on a sidechain used for managing other sidechains. The Management and Pinning contract specifies who are the participants of the sidechain and holds information which pins the state of the sidechain to contract. 

For each sidechain, there are masked and unmasked participants. Unmasked Participants have their Ethereum Addresses listed as being members of the sidechain. Being unmasked allows the participant to vote to add and remove other participants, to change the voting period and algorithm, and to contest pins.

Masked Participants are participants that are listed against a sidechain in a way that observers can not determine their identity. These participants are represented as a salted hash of their Ethereum Address. The masked participant keeps the salt secret value off-chain. If they need to unmask themselves, they present their secret salt to the contract. This combined with their sending address is used to create the calculated salted hash. If this calculated value matches their masked participant value then they become an unmasked participant. The method of masking participants allows the solution to comply with the requirement SC-4b-PinningParticipantShielding.

Pinning values are put into a map. All participants of a sidechain agree on a Sidechain Secret. The Sidechain Secret seeds a Pseudo Random Number Generator (PRNG). A new 256 bit value is generated each time an uncontested pin is posted. The key in the map is calculated using the equations shown below.

\begin{equation}
PrngValue = PRNG.nextValue
\end{equation}
\begin{multline}
MapKey = Keccak256( Sidechain Identifier, \\
Previous Pin, PrngValue)
\end{multline}

Where the initial value of Previous Pin is zero. The PRNG algorithm needs to be agreed upon by the sidechain participants when the sidechain is being created. 

Masked and unmasked participants of a sidechain observe the pinning map at the MapKey value waiting for the next pin to be posted to that entry in the map. When the pin value is posted, they can then determine if they wish to contest the pin. To contest the pin, they submit to the contract:

\begin{itemize}
\item Previous MapKey.
\item PrngValue.
\item Sidechain Identifier.
\end{itemize}

Submitting the previous value of the MapKey allows the contract to fetch from its own storage the value of the Previous Pin. The contract can then calculate the MapKey of the contested pin by combining the Previous Pin, PrngValue and Sidechain Identifier using the equation above. Given the submitter of the transaction knows PrngValue which combined with the Sidechain Identifier links the Previous MapKey and the calculated MapKey, it implies that the MapKeys correspond to pins for the sidechain. The further implication of knowing a value PrngValue is that the transaction submitter has access to the Sidechain Seed and are a member of the sidechain.

Once a pin is marked as contested, the unmasked participants of the sidechain must then vote on the validity of the posted pin. At this point, masked participants will need to unmask themselves to vote. This could be viewed as analogous to a situation in which companies that have a traditional private written agreement are in dispute. Under normal circumstances the companies may be able to keep their agreement private. However, in case of dispute, the companies need to make public their agreement and go to the courts to resolve the matter. Similarly, masked participants need to unmask themselves if they wish to vote on a disputed pin. This methodology for contesting pins allows this solution to comply with requirement SC-4d-PinningContesting.

Multiple sidechains could be managed from the one Management and Pinning contract. This would be advantageous because it would mean that the pin values for the different sidechains could be put intermixed in the one pinning map. This would hide the rate of pinning for any one sidechain. This allows the solution to comply with SC-4c-PinningTransactionRateShielding.

To meet requirement SC-4e-PinningCipherTextObservers, sidechain participants may choose to use a Quiet Guardian. This is an entity which could be the only unmasked participant of a sidechain. It could be configured to only see the encrypted transactions and state of the sidechain, and post that encrypted state as pins to the Management and Pinning contract. Having a Quiet Guardian as the lone unmasked participant of a sidechain allows the other sidechain participants to remain anonymous, assuming no dispute.

\subsection{Interchain Communications}
Three types of inter-sidechain and inter-blockchain communications are possible: reading, writing and transferring value. Reading means executing a Solidity View function which does not update the blockchain state. Writing means executing function calls which result in transactions which require updates to the blockchain. Value transfer relates to transferring Ether or an ERC20 token \cite{eip20}\cite{erc20-standard}.

Any inter-sidechain communications requires correct permissions, such that an entity can read form or write to the target sidechain. In the case of a permissionless sidechain, this would allow any entity to read from or write to it. In the case of a permissioned sidechain reading or writing to another permissioned sidechain, the participant's Ethereum Address used on each sidechain needs to be the same.

When a participant that is using one sidechain wishes to perform a read from a target sidechain, they need to ensure that the block number from the target sidechain is referenced in the transaction. Doing this allows all nodes which execute the transaction on the originating sidechain read from the same context on the target sidechain. Nodes on the originating sidechain need to assess whether they perceive the block number on the target sidechain is recent before accepting the transaction.

Inter-sidechain transactions which cause state updates on a target sidechain is an area which needs further work. A possible solution is to have each node on the originating sidechain reference the same block of the target sidechain. They can each perform a trial execution of the transaction to see that they come to the same value.

Hashed Timelock Contracts \cite{hashtimelock} have been put forward for inter-chain value transfer. It could be envisaged that a participant could choose to put some Ether in escrow in a contract on Ethereum MainNet. They could launch a sidechain which issued them the same amount of Ether previously escrowed. They could then use the concept of Hashed Timelock Contracts to allow trustless transfer of Ether between Ethereum MainNet and the sidechain. Given a fixed total quantity of Ether on the sidechain, this would allow for ``Mass Conservation" between Ethereum MainNet and the sidechain, where no additional Ether is created or destroyed. 

Building on the concepts of Hashed Timelock Contracts, Thomas and Schwartz have proposed an Interledger Protocol \cite{interledger}. Either this technology or other similar technologies could be used to allow inter-blockchain value transfers.

\subsection{Sidechain Capable Ethereum Client Architecture}
\begin{figure}
  \includegraphics[width=\linewidth]{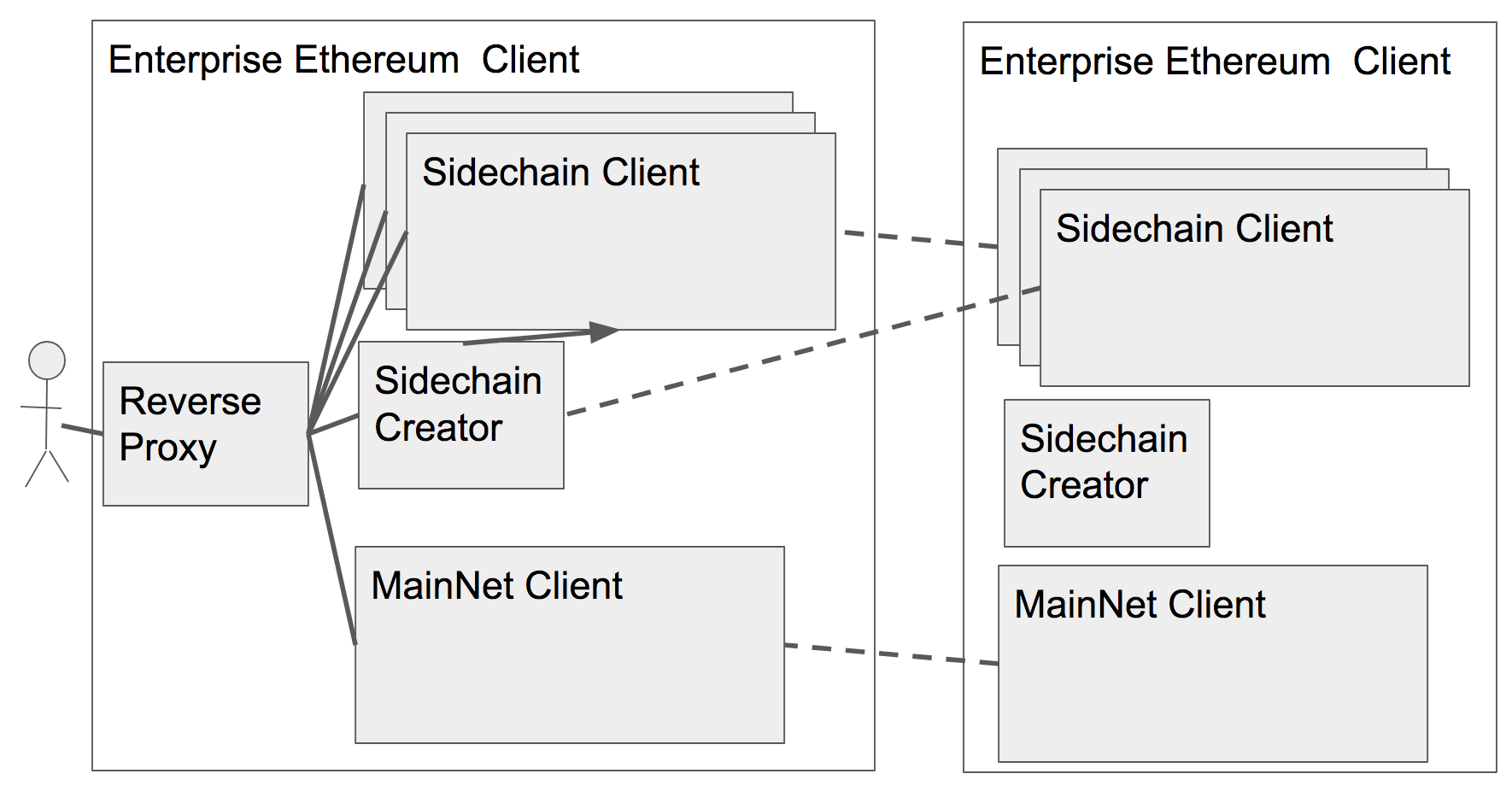}
  \caption{Sidechain Client Architecture}
  \label{fig:sidechain-client-architecture}
\end{figure}

Figure \ref{fig:sidechain-client-architecture} shows a simplified architecture of an Enterprise Ethereum Client which includes sidechain capability, showing only JSON RPC calls and some network connections. 

JSON RPC requests come from DApps represented in the diagram by the stick figure. These go via a reverse proxy which splits the traffic based on Sidechain Identifier. For requests with no Sidechain Identifier, the requests are routed to the MainNet Client. For requests with a Sidechain Identifier, requests are routed to the sidechain with the specified Sidechain Identifier. Special requests to create sidechains are routed to the Sidechain Creator. 

When the Sidechain Creator receives a request over the JSON RPC to create a sidechain, it checks that the requesting participant has been authorised to create sidechains, it gathers bootstrap information from Ethereum Registration Authorities to determine which nodes to connect establish connections with, and instantiates a Sidechain Client. The Sidechain Client contacts other Enterprise Ethereum Clients' Sidechain Creators requesting they join the sidechain. These remote Sidechain Creators check whether the requesting node belongs to an organisation which is either whitelisted or not blacklisted. Assuming the incoming request is from an authorised organisation, the Sidechain Creator instantiates a Sidechain Client which works with the remote Sidechain Client to establish the sidechain network.

All APIs of the Enterprise Ethereum Client need to be permissioned, ensuring that only authorised participants can use them. To allow standard Ethereum APIs to be used, an appropriately permissioned participant could interact directly with the Sidechain Client API.

\subsection{Adding and Removing Nodes and Organisations}
Only nodes belonging to organisations which are a part of a sidechain are authorised to communicate with other nodes on a sidechain. The list of nodes belonging to an organisation can be found by looking up information from the ERA. When an organisation wishes to add a node, it updates its OrgInfo information, and then sends an indication to all other organisations on the network. The other organisations which are part of the sidechain can check the ERAs, and establish communications with the new node. 

To add an organisation to or remove an organisation from a sidechain, unmasked participants for the sidechain in a Management and Pinning contract propose and then vote on adding or removing the organisation. Once the organisation has been added, the ERA can be used to determine the nodes to be added or removed from the sidechain network.

\subsection{Encrypted State and Encrypted Transactions}
The state of each sidechain needs to be authenticated encrypted using separate keys for each sidechain. This ensures that sidechain state from different sidechains are mutually encrypted. The reason why this is important is that in the future, rather that individual companies operating Ethereum nodes or sidechain nodes, a ``sidechain hosting company" might provides services to host the Sidechain Clients for an individual. In a similar way to how cloud tenants expect their data to be encrypted with a different key to other tenants, users of sidechains will expect their data to be separately encrypted. This one key per sidechain approach could be achieved by having a root key and deriving the key per sidechain from the root key based on the Sidechain Identifier. Alternatively, each sidechain key could be randomly generated, and a key manager could be used to store sidechain keys based on Sidechain Identifier.

\subsection{Hyperledger Fabric Sidechains}
Hyperledger Fabric's Channels meet many of the requirements of Ethereum Private Sidechains. Adding support for a Byzantine Fault Tolerant consensus algorithm, and encrypting the state of channels are important changes which should be made to improve the security of the platform. In a private conversation with John Wolpert, he suggested that Hyperledger Fabric could benefit from supporting the use of ERAs and Management and Pinning contracts. Using ERAs would allow Ethereum MainNet or an Ethereum sidechain to used as a way for entities which have not previously interacted to set-up a network and channel together. Using Management and Pinning contracts would allow Hyperledger Fabric to leverage Ethereum MainNet's improved security whilst maintaining privacy.

\section{Future Work}
A high level overview of the Ethereum Private Sidechains system has been provided in this paper. A more detailed exploration and analysis of the components of this system should be undertaken. Additionally, more work should be undertaken in the following areas:

\begin{itemize}
\item Inter-sidechain communications mechanisms, with a focus on inter-chain transactions.
\item Organisationally aware consensus and voting algorithms.
\item Quantifiable performance requirements that match the expectations of users of Enterprise Ethereum Clients.
\end{itemize}

The requirements in this paper were measured against three blockchain platforms. These requirements should be analysed against other platforms such as R3 Corda \cite{corda-doc}.

\section{Conclusion}
Using the Enterprise Ethereum Client Specification as a basis, this paper has defined a set of requirements which Enterprise Ethereum Clients that wish to offer private sidechains capabilities should offer. None of the three blockchain clients analysed met all of the requirements, all having their strengths and weaknesses. Quorum supports IBFT, a Byzantine Fault Tolerant algorithms, whereas Parity's Private Transactions require all validators to approve the transaction, and Hyperledger Fabric's Kafka relies on no parties acting maliciously. Parity has encrypted state whereas Quorum and Hyperledger Fabric store state in plaintext. Hyperledger Fabric supports channels, which are similar to sidechain and has the concept of organisation, whilst Quorum and Parity do not support sidechains and don't have the concept or organisations.

The architecture of a Ethereum Private Sidechains system has been presented. Included in this architecture is how this system relates to Ethereum MainNet and how to incorporate other sidechain like systems such as Hyperledger Fabric. The proposed architecture includes the Ethereum Registration Authority system which allows for information to securely placed onto the Ethereum blockchain such that it is discoverable and describes a Management and Pinning contract which pins the state of a sidechain to Ethereum MainNet without revealing the identities of sidechain participants and whilst protecting the transaction rate of the sidechain.

\begin{appendices}
\section{Enterprise Ethereum Client Specification Requirements}
\subsection{Introduction}
The requirements shown in this appendix are requirements listed in the Enterprise Ethereum Client Specification 1.0. All requirements are denoted with a label. The label starts with ``EE" to indicate an Enterprise Ethereum Client Specification based requirement. This is followed by the section number from the specification, which is followed by a letter to indicate the sequence number of the requirement within the section. The label is finished with a brief summary of the requirement.

\subsection{Application Layer}
\subsubsection{DApps Sublayer}
\begin{itemize}
\item EE-4.1a-DApp: ``ÐApps MAY use the Enterprise Ethereum extension to the JSON-RPC API defined in this Specification."
\end{itemize}

\subsubsection{Smart Contract Tools Sublayer}
\begin{itemize}
\item EE-4.3a-Tools: ``Implementations MUST provide deployment and debugging tools for Enterprise Ethereum smart contracts."
\item EE-4.3b-FormalVerification: ``Implementations SHOULD extend formal verification methods for use with Enterprise Ethereum smart contracts."
\end{itemize}

\subsection{Tooling Layer}
\subsubsection{Permissioning and Credentials Sublayer}
.

\underline{Nodes}

\begin{itemize}
\item EE-5.1.1a-StaticStartUp: ``Enterprise Ethereum implementations MUST provide the ability to specify at startup a list of static peer nodes to establish peer-to-peer connections with."
\item EE-5.1.1b-DisableDiscovery: ``Implementations MUST provide the ability to enable or disable peer-to-peer node discovery."
\item EE-5.1.1c-WhitelistNodes: ``Implementations MUST provide the ability to specify a whitelist of the node identities permitted to join the network."
\item EE-5.1.1d-BlacklistNodes: ``Implementations MAY provide the ability to specify a blacklist of the node identities not permitted to join the network."
\item EE-5.1.1e-WhitelistViaAPI: ``It MUST be possible to specify the node whitelist through an interface or API."
\item EE-5.1.1f-BlacklistViaAPI: ``It MUST be possible to specify the node blacklist (if implemented) through an interface or API."
\item EE-5.1.1g-CertifyNodes: ``Implementations MUST provide a way to certify the identities of nodes."
\item EE-5.1.1h-Organization: ``An Enterprise Ethereum client SHOULD provide mechanisms to define clusters of nodes at the organizational level, in the context of permissioning."
\end{itemize}

\underline{Participants}
\begin{itemize}
\item EE-5.1.2a-WhitelistParticipants: ``Implementations MUST provide the ability to specify a whitelist of participant identities who are permitted to submit transactions."
\item EE-5.1.2b-BlacklistParticipants: ``Implementations MAY provide the ability to specify a blacklist of participant identities who are not permitted to submit transactions."
\item EE-5.1.2c-WhitelistParticipantsViaAPI: ``It MUST be possible to specify the participant whitelist through an interface or API."
\item EE-5.1.2d-BlacklistParticipantsViaAPI: ``It MUST be possible to specify the participant blacklist (if implemented) through an interface or API."
\item EE-5.1.2e-CertifyParticipants: ``Implementations MUST provide a way to certify the identities of participants."
\item EE-5.1.2f-GroupsRoles: ``Implementations MUST provide the ability to specify participant identities in a way aligned with the usual concepts of groups and roles."
\end{itemize}

\underline{Additional Requirements}
\begin{itemize}
\item EE-5.1.3a-SmartContractPermissioning: ``Implementations SHOULD provide permissioning schemes through standard mechanisms, such as smart contracts used in a modular way. That is, permissioning schemes could be implemented to interact with smart contract-based mechanisms."
\item EE-5.1.3b-RuntimeConfigUpdate: ``Implementations SHOULD provide the ability for configuration to be updated at run time without the need to restart."
\item EE-5.1.3c-ConfigOptions: ``Implementations MAY provide configuration through the use of flat files, command-line options, or configuration management system interfaces."
\item EE-5.1.3d-LocalKeyManagement: ``Implementations MAY support local key management allowing users to secure their private keys."
\item EE-5.1.3e-SecureExternalKeyGenStore: ``Implementations MAY support secure interaction with an external Key Management System for key generation and secure key storage."
\item EE-5.1.3f-HardwareSecurityModules: ``Implementations MAY support secure interaction with a Hardware Security Module (HSM) for deployments where higher security levels are needed."
\end{itemize}

\subsubsection{Integration and Deployment Sublayer}
.
\underline{Integration Libraries}
\begin{itemize}
\item EE-5.2.1a-IntegrationLibraries: ``Implementations MAY provide integration libraries enabling convenience of interaction through additional language bindings."
\end{itemize}

\underline{Enterprise Management Systems}
\begin{itemize}
\item EE-5.2.2a-EntDeployment: ``Implementations SHOULD provide enterprise-ready software deployment and configuration capabilities, including the ability to easily: Deploy through enterprise remote software deployment and configuration systems; Modify configurations on already deployed systems;
Audit configurations on already deployed systems."
\item EE-5.2.2b-EntFaultReporting: ``Implementations SHOULD provide enterprise-ready software fault reporting capabilities, including the ability to: Log software fault conditions; Generate events to notify of software fault conditions; Accept diagnostic commands from software fault management systems."
\item EE-5.2.2c-EntPerformanceManage: ``Implementations MAY provide enterprise-ready performance management capabilities, including the ability to easily provide relevant performance management metrics for analysis by enterprise performance management systems."
\item EE-5.2.2d-EntSecurity: ``Implementations SHOULD provide enterprise-ready security management interaction capabilities, including the ability for: Logs to be easily monitored by enterprise security management systems; Events to be easily monitored by enterprise security management systems; Secure network traffic to be monitored by enterprise security management systems."
\item EE-5.2.2e-EntHistoricalAnalysis: ``Implementations MAY provide enterprise-ready capabilities to support historical analysis, including the ability for relevant metrics to be easily collected by an enterprise data warehouse system for detailed historical analysis and creating analytical reports."
\item EE-5.2.2f-EntManagementSystems: ``Implementations MAY include support for other enterprise management systems, as appropriate, such as: Common Management Information Protocol (CMIP), Web-Based Enterprise Management (WBEM), Application Service Management (ASM) instrumentation."
\end{itemize}

\subsubsection{Client Interfaces Sublayer}
.

\underline{Extensions to the JSON-RPC API}
\begin{itemize}
\item EE-5.3.1a-JsonRpcPublicEth: ``Implementations MUST provide support for the public Ethereum JSON-RPC API."
\item EE-5.3.1b-JsonRpcTransactionAsyncExt: ``Implementations MUST provide the eth\_sendTransactionAsync Enterprise Ethereum extension call to the public Ethereum [JSON-RPC API] for at least one of the private transaction types defined in Section 6.1.2."
\item EE-5.3.1c-JsonRpcUnimplemented: ``The eth\_sendTransactionAsync call MUST respond with an HTTP 501 (Not Implemented) status code when an unimplemented private transaction type is requested."
\end{itemize}

\underline{Inter-Chain}
\begin{itemize}
\item EE-5.3.2a-InterChainInteraction: ``Enterprise Ethereum implementations MAY provide inter-chain mediation capabilities to enable interaction with different blockchains."
\end{itemize}

\underline{Oracles}
\begin{itemize}
\item EE-5.3.3a-Oracles: ``Enterprise Ethereum implementations SHOULD provide the ability to securely interact with oracles to send and receive real-world information."
\end{itemize}

\subsection{Privacy and Scaling Layer}
\subsubsection{Privacy Sublayer}
.

\underline{On-Chain}
\begin{itemize}
\item EE-6.1.1a-OnChainSecurity: ``Implementations SHOULD support improved on-chain security techniques as they become available."
\end{itemize}

\underline{Private Transactions}
\begin{itemize}
\item EE-6.1.2a-PrivTransMethods: ``Implementations MUST support private transactions using at least one of the following methods: Private transactions where payload data is transmitted to and readable only by the direct participants of a transaction. These transactions are referred to as restricted private transactions;
Private transactions where payload data is transmitted to all nodes participating in the network but readable only by the direct participants of a transaction. These transactions are referred to as unrestricted private transactions."
\end{itemize}

\underline{Restricted Private Transactions}
\begin{itemize}
\item EE-6.1.2b-RestrictedPayloadMaskingStored: ``Implementations MUST support masking or obfuscation of the payload data when stored in restricted private transactions (for example, using cryptographic encryption)."
\item EE-6.1.2c-RestrictedPayloadMaskingTransit: ``Implementations MUST support masking or obfuscation of the payload data when in transit in restricted private transactions (for example, using cryptographic encryption)."
\item EE-6.1.2d-RestrictedMetadataMaskingStored: ``Implementations MAY support masking or obfuscation of the metadata when stored in restricted private  transactions (for example, using cryptographic encryption)."
\item EE-6.1.2e-RestrictedMetadataMaskingTransit: ``Implementations MAY support masking or obfuscation of the metadata when in transit in restricted private transactions (for example, using cryptographic encryption)."
\item EE-6.1.2f-RestrictedPayloadRelayStore: ``Nodes that relay a restricted private transaction but are not participants in that transaction MUST NOT store transaction payload data."
\item EE-6.1.2g-RestrictedMetadataRelayStore: ``Nodes that relay a restricted private transaction but are not participants in that transaction SHOULD NOT store metadata."
\item EE-6.1.2h-RestrictedDefaultSecure: ``The implementation of the JSON RPC API eth\_sendTransactionAsync call (if implemented), either without the restriction parameter or with the restriction parameter set to restricted, MUST result in a restricted private transaction."
\end{itemize}

\underline{Unrestricted Private Transactions}
\begin{itemize}
\item EE-6.1.2i-UnrestrictedRecipientMasking: ``Implementations SHOULD support masking or obfuscation of the recipient identity when stored in unrestricted private transactions (for example, using cryptographic encryption, or ring signatures and mixing)."
\item EE-6.1.2j-UnrestrictedSenderMasking: ``Implementations SHOULD support masking or obfuscation of the sender identity when stored in unrestricted private transactions (for example, using stealth addresses)."
\item EE-6.1.2k-UnrestrictedPayloadMaskingStored: ``Implementations SHOULD support masking or obfuscation of the payload data when stored in unrestricted private transactions (for example, using cryptographic encryption)."
\item EE-6.1.2l-UnrestrictedPayloadMaskingTransit: ``Implementations MUST support masking or obfuscation of the payload data when in transit in unrestricted private transactions (for example, using cryptographic encryption)."
\item EE-6.1.2m-UnrestrictedMetadataMaskingStored: ``Implementations MAY support masking or obfuscation of the metadata when stored in unrestricted private transactions (for example, using cryptographic encryption)."
\item EE-6.1.2n-UnrestrictedMetadataMaskingTransit: ``Implementations MAY support masking or obfuscation of the metadata when in transit in unrestricted private transactions (for example, using cryptographic encryption)."
\item EE-6.1.2o-UnrestrictedPayloadRelayStore: ``Nodes that relay an unrestricted private transaction but are not participants in that transaction MAY store payload data."
\item EE-6.1.2p-UnrestrictedMetadataRelayStore: ``Nodes that relay an unrestricted private transaction but are not participants in that transaction MAY store transaction metadata."
\item EE-6.1.2q-UnrestrictedTransactions: ``The implementation of the JSON RPC API eth\_sendTransactionAsync call (if implemented) with the restriction parameter set to unrestricted MUST result in an unrestricted private transaction."
\end{itemize}

\underline{Private Transactions Other}
\begin{itemize}
\item EE-6.1.2r-PrivateTransactionAddParticipants: ``Implementations SHOULD be able to extend the set of participants in a private transaction (or forward the private transaction in some way)."
\item EE-6.1.2s-PrivateTransactionConsensus: ``Implementations SHOULD provide the ability for nodes to achieve consensus on their mutually private transactions."
\end{itemize}

\subsubsection{Scaling Sublayer}
.

\underline{Off-Chain (Trusted Execution)}
\begin{itemize}
\item EE-6.1.3a-OffchainTrustedExecution: ``Enterprise Ethereum implementations SHOULD provide the ability for off-chain, trusted execution of transactions and smart contracts."
\end{itemize}

\underline{On-Chain (Layer 2)}
\begin{itemize}
\item EE-6.2.1a-ImprovedOnchainProcessing: ``Enterprise Ethereum implementations SHOULD provide the ability for improved on-chain processing rates of transactions and smart contracts."
\end{itemize}

\underline{Off-Chain (Compute)}
\begin{itemize}
\item EE-6.2.2a-OffchainProcessing: ``Enterprise Ethereum implementations SHOULD provide the ability for off-chain processing of transactions and smart contracts."
\end{itemize}

\underline{Performance}
\begin{itemize}
\item EE-6.2.3a-ArchivePrivateState: ``Implementations SHOULD support the ability to have private state data archived from the blockchain while preserving the consistency and validity of the blockchain."
\item EE-6.2.3b-ComputePowerSizeParticipants: ``The computing power to validate blocks SHOULD remain constant over time, regardless of the blockchain size or the number of network participants."
\item EE-6.2.3c-RecentBlockAccessTime: ``The time to access recent blockchain data SHOULD remain constant, regardless of the blockchain size."
\item EE-6.2.3d-NewGenesisBlock: ``Implementations SHOULD allow network operators to designate a new genesis block to keep the blockchain size from growing perpetually."
\end{itemize}

\subsection{Core Blockchain Layer}
\subsubsection{Storage and Ledger Sublayer}
\begin{itemize}
\item EE-7.1a-StoragePubEth: ``Enterprise Ethereum implementations SHOULD implement data storage requirements necessary to operate a public Ethereum client."
\item EE-7.1b-StorageForOptionalOffchain: ``Implementations MAY implement data storage used for optional off-chain operations."
\item EE-7.1c-SeparateStoragePerNetwork: ``Implementations providing support for multiple networks (for example, one or more consortium networks or a public network) MUST store data related to private transactions for those networks in private state dedicated to the relevant network."
\item EE-7.1d-DataAccessSameParticipants: ``A smart contract operating on private state SHOULD be permitted to access private state created by other smart contracts involving the same participants."
\item EE-7.1e-DataAccessDifferentParticipants: ``A smart contract operating on private state MUST NOT be permitted to access private state created by other smart contracts involving different participants."
\item EE-7.1f-FileDecentralizedStorage: ``Implementations SHOULD provide the ability for private smart contracts to store file objects seamlessly and transparently, so no artificial off-chain file-storage add-ons are needed."
\end{itemize}

\subsubsection{Execution Sublayer}
\begin{itemize}
\item EE-7.2a-EvmOpCodes: ``Enterprise Ethereum implementations MUST provide a smart contract execution environment implementing the public Ethereum EVM op-code set [EVM Opcodes]."
\item EE-7.2b-EvmExtendedOpCodes: ``Enterprise Ethereum implementations MAY provide a smart contract execution environment extending the public Ethereum EVM op-code set [EVM Opcodes]."
\item EE-7.2c-PublicStateSync: ``Implementations SHOULD support the ability to synchronize their public state with the public state held by other public Ethereum nodes."
\item EE-7.2d-PrecompiledContracts: ``Implementations MAY provide support for the compilation, storage, and execution of precompiled contracts."
\item EE-7.2e-TEE: ``TEEs ensure only authorized parties can execute smart contracts on an execution environment related to a given consortium network. Implementations SHOULD provide a TEE."
\item EE-7.2f-TEEConfigurableEncryption: ``Multiple encryption techniques could be used to secure TEEs or private state. Implementations SHOULD provide configurable encryption options for use in conjunction with consortium networks."
\item EE-7.2.1a-Finality: ``When a deterministic consensus algorithm is used, transactions SHOULD be considered final after a defined interval or event. This interval may be a set time period or a set number of blocks being created since the transaction was included in a block."
\end{itemize}

\subsubsection{Consensus Sublayer}
\begin{itemize}
\item EE-7.3a-MainNetConsensus: ``Enterprise Ethereum implementations SHOULD support the ability to form consensus on Ethereum MainNet (public Ethereum) and to form consensus operating as part of an Enterprise Ethereum network."
\item EE-7.3b-MultipleConsensusAlgorithms: ``Implementations MUST be capable of supporting multiple consensus algorithms."
\item EE-7.3c-PrivateConsensusAlg: ``One or more consensus algorithms SHOULD allow operations as part of an Enterprise Ethereum network."
\item EE-7.3d-MainNetConsensusAlg: ``One or more consensus algorithms SHOULD allow operations on the Ethereum MainNet.
\item EE-7.3e-SidechainConsensusAlg: ``One or more consensus algorithms MAY support operations on sidechain networks."
\item EE-7.3f-ConsensusAlgDocumented: ``Consensus algorithms MUST be clearly documented for interoperability."
\item EE-7.3g-ConsensusAlgModularConf: ``Consensus algorithm implementations SHOULD be modular and configurable."
\item EE-7.3h-ConsensusInOutOfBand: ``Consensus algorithms MAY communicate in-band or out-of-band with other clients, as requested. That is, consensus algorithm implementations can make and receive network traffic external to the client-to-client network protocol."
\item EE-7.3i-ConsensusIBFT: ``Implementations SHOULD support the Istanbul [Byzantine Fault Tolerance] (IBFT) consensus algorithm [EIP-650], so individual attacked or malfunctioning clients performing voting, block-making, or validation roles do not pose a critical risk to the network."
\item EE-7.3j-ConsensusOther: ``Implementations MAY support other consensus algorithms."
\item EE-7.3k-ConsensusConfig: ``Implementations MUST provide the ability to specify the consensus algorithms, through configuration, to be used for each public blockchain, private blockchain and sidechain in use."
\end{itemize}

\subsection{Network Layer}
\subsubsection{Network Protocol Sublayer}
\begin{itemize}
\item EE-8.1a-Enode: ``Nodes MUST be identified and advertised using the Ethereum enode URL format [enode]."
\item EE-8.1b-DevP2P: ``Implementations SHOULD use the DEVp2p Wire Protocol [DEVp2p Wire Protocol] for messaging between nodes to establish and maintain a communications channel for use by higher layer protocols."
\item EE-8.1c-Eth62Eth63: ``Implementations SHOULD support, at a minimum, [Ethereum Wire Protocols] eth/62 and eth/63."
\item EE-8.1d-NewProtocols: ``Implementations MAY add new protocols or extend existing Ethereum protocols."
\item EE-8.1e-RelayNodes: ``To minimize the number of point-to-point connections needed between private nodes, some private nodes SHOULD be capable of relaying private transaction data to multiple other private nodes."
\end{itemize}

\subsubsection{Anti-Spam}
\begin{itemize}
\item EE-9a-AntiSpam: ``Enterprise Ethereum implementations SHOULD provide effective anti-spam mechanisms so attacking nodes or addresses (either malicious, buggy, or uncontrolled) can be quickly identified and stopped."
\end{itemize}

\subsection{Cross-client Compatibility}
\begin{itemize}
\item EE-10a-PublicEthCompatibility: ``Enterprise Ethereum clients SHOULD be compatible with the public Ethereum network to the greatest extent possible."
\item EE-10b-ExtendedApisSuperset: ``Implementations MAY extend the public Ethereum APIs. To maintain compatibility, implementations SHOULD ensure these new features are a superset of the public Ethereum APIs."
\end{itemize}

\subsection{Synchronization and Disaster Recovery}
\begin{itemize}
\item EE-11a-FastSync: ``Implementations SHOULD support a fast synchronization mode so new clients can be launched quickly and synchronized to long standing, historical blockchains with the understanding that the new client might not have the complete blockchain history."
\item EE-11b-BackupRestore: ``Implementations SHOULD support a mechanism to back up data and use it later to initialize a node, up to a certain block."
\end{itemize}

\section{Glossary}
\label{glossary}
Table \ref{table_glossary} presents a list of terms used in this paper.

\begin{table*}[t]
  \centering
    \begin{tabular}{| l | l |}
    \hline
    Term                  & Definition \\
    \hline
    Access Control  & Checks that an authenticated identity is authorised to use a resource.     \\
    \hline
    Authentication   & Confirms the claimed identity of a party interacting with a service.    \\
    \hline
    Authorisation     & Confers a permission to perform an action by an authenticated identity on a resource.    \\
    \hline
    Chaincode        & The Hyperledger Fabric term for a Smart Contract.    \\
    \hline
    Client                 & The software used to operate a blockchain node.    \\
    \hline
    Contract            & Another term for a Smart Contract.    \\
    \hline
    MainNet            & The Ethereum public chain.     \\
    \hline
    Management Chain & A blockchain or sidechain which is used to coordinate the activities of a private sidechain.     \\
    \hline
    Network Participant & Another term for node. In particular, this is indicating the node is an entity on a peer-to-peer network taking part in the     \\
                              & distributed system.    \\
    \hline
    Node                 & An instance of blockchain client software operating as part of a peer-to-peer blockchain network.    \\
    \hline
    Oracle.              & A service external to the blockchain that is trusted by the creators of smart contracts and is called to provide information.\\
                              & For example, an Oracle could be a service to return a current exchange rate.\\
    \hline
    Participant         & A user of the API of a blockchain node.  The user may be a human user or an automated computer users.    \\
    \hline
    Peer                  & The term used in Hyperledger Fabric for Node.    \\
    \hline
    Permissioned    & A blockchain, sidechain, or smart contract which performs access control to ensure only authenticated, authorised users \\
                              & can perform actions on the resource.    \\
    \hline
    Permissionless  & A blockchain, sidechain, or smart contract which is open to all participants. That is, no access control or authentication is  \\
                              & performed as all participants can use the resource. Unpermissioned and non-permissioned are terms used at times in \\
                              & literature to refer to permissionless resources.   \\
    \hline
    Pinning              & Pinning, sometimes known as Pegging, is the act of posting a summary of the state of a sidechain to a management chain.      \\
    \hline
    Private               & A blockchain, sidechain, or smart contract which is permissioned. That is, authentication, authorisation, and access control \\
                              & are performed on the blockchain, sidechain, or smart contract.      \\
    \hline
    Private Sidechain & A permissioned blockchain which hosts private smart contracts. Ethereum Private Sidechains offer permissioning such that \\
                              & only nodes which belong to organisations which are a party the sidechain have access to the sidechain transactions or state. \\
                              & The Ethereum Sidechain Client APIs are permissioned such that only authorised users can interact with them, and only \\
                              & authorised Ethereum Accounts can be used to submit transactions form them. The transactions and the state of sidechains \\
                              &  are authenticated-encrypted to ensure data confidentiality. \\
    \hline
    Public                 & A permissionless blockchain, sidechain, or smart contract. That is, no authentication, authorisation, or access control \\
                              & are performed to gain access to the blockchain, sidechain, or smart contract.      \\
    \hline
    Sidechain          & A permissioned or permissionless blockchain which is managed via a management chain. \\
    \hline
    Smart Contract  & A computer program uploaded onto the blockchain.    \\
    \hline

  \end{tabular}
  \caption{Glossary}
  \label{table_glossary}
\end{table*}

\end{appendices}

\begin{table*}[t]
  \centering
  \begin{tabular}{|| l || c | c | c || *{6}{c |}}
    \hline
    Requirements                       & MUST or     & SHOULD or     & MAY               & Quorum          & Parity's                & Hyperledger     \\
                                                & MUST NOT & SHOULD &                        &                        & Private                         & Fabric                           \\
                                                &                     & NOT        &                        &                        & Transactions                &                            \\
    \hline
    EE-4.3a-Tools                      & $\checkmark$ &                      &                    &  $\checkmark$ & $\times$                  & $\checkmark$  \\
    \hline
    EE-5.1.1a-StaticStartUp      & $\checkmark$ &                      &                     & $\checkmark$ & $\checkmark$          & $\checkmark$ \\
    \hline
    EE-5.1.1b-DisableDiscovery & $\checkmark$ &                      &                    & $\times$          & N/A                          & $\times$          \\
    \hline
    EE-5.1.1c-WhitelistNodes    & $\checkmark$ &                      &                    & $\checkmark$  & $\checkmark$         & $\checkmark$ \\
    \hline
    EE-5.1.1e-WhitelistViaAPI    & $\checkmark$ &                      &                    &  $\times$           & $\checkmark$          & $\checkmark$ \\
    \hline
    EE-5.1.1f-BlacklistViaAPI     & $\checkmark$ &                      &                    &  N/A                   & N/A                         & N/A                 \\
    \hline
    EE-5.1.1h-CertifyNodes        & $\checkmark$ &                      &                    &  $\times$           & $\times$                  & $\checkmark$ \\
    \hline
    EE-5.1.2a-WhitelistParticipants & $\checkmark$ &                  &                    & $\times$            & $\times$                  & $\checkmark$ \\
    \hline
    EE-5.1.2c-WhitelistParticipantsViaAPI & $\checkmark$ &         &                    & $\times$          & $\times$                 &  $\checkmark$ \\
    \hline
    EE-5.1.2d-BlacklistParticipantsViaAPI & $\checkmark$ &          &                   & N/A                  & N/A                         &  N/A                 \\
    \hline
    EE-5.1.2e-CertifyParticipants    & $\checkmark$ &                   &                    & $\times$           &  $\times$                 &  $\checkmark$ \\
    \hline
    EE-5.1.2f-GroupsRoles      & $\checkmark$ &                          &                    & $\times$           &  $\times$                 &  $\checkmark$ \\
    \hline
    EE-5.3.1a-JsonRpcJsonRpcPublicEth & $\checkmark$ &         &                   & $\checkmark$   & $\checkmark$         &  Ethereum      \\
    \hline
    EE-5.3.1b-JsonRpcTransactionAsyncExt & $\checkmark$ &      &                  & $\times$            & $\times$                &   Ethereum      \\
    \hline
    EE-5.3.1c-JsonRpcUnimplemented   & $\checkmark$ &              &                 & $\times$           & $\times$                 &   Ethereum       \\
        \hline
    EE-6.1.2a-PrivTransMethods   & $\checkmark$ &                          &               & $\checkmark$ &  $\checkmark$        & $\checkmark$   \\
    \hline
    EE-6.1.2b-RestrictedPayloadMaskingStored  & $\checkmark$ &             &          & $\checkmark$ With caveats. &  N/A & $\checkmark$  \\
                                                                            &                         &             &          & See section \ref{EE-6.1.2b-RestrictedPayloadMaskingStored} &        &    \\
                                                                            &                         &             &          & for details. &        &   \\
    \hline
    EE-6.1.2c-RestrictedPayloadMaskingTransit & $\checkmark$ &              &          & $\checkmark$  &  $\checkmark$   & $\checkmark$   \\
    \hline
    EE-6.1.2f-RestrictedPayloadRelayStore       & $\checkmark$ &              &          & N/A                     & N/A                       & N/A                 \\
    \hline
    EE-6.1.2h-RestrictedDefaultSecure                 & $\checkmark$ &              &          & N/A                  & N/A                     & Ethereum         \\
    \hline
    EE-6.1.2l-UnrestrictedPayloadMaskingTransit  & $\checkmark$ &           &           & N/A                  &  N/A                    & N/A                   \\
    \hline
    EE-6.1.2q-UnrestrictedTransactions                 & $\checkmark$ &            &             & N/A                & N/A                   & N/A                   \\
    \hline
    EE-7.1c-SeparateStoragePerNetwork                & $\checkmark$ &              &          & N/A                & $\times$           &  $\times$          \\
    \hline
    EE-7.1e-DataAccessDifferentParticipants          & $\checkmark$ &              &          & $\times$           & $\checkmark$ &  $\times$            \\
    \hline
    EE-7.2a-EvmOpCodes                                       & $\checkmark$ &              &          &  $\checkmark$  & $\checkmark$ & Ethereum      \\
    \hline
    EE-7.3b-MultipleConsensusAlgorithms               & $\checkmark$ &              &          & $\checkmark$  & $\times$          &  $\checkmark$  \\
    \hline
    EE-7.3f-ConsensusAlgDocumented                    & $\checkmark$ &              &          & $\checkmark$  & $\checkmark$ & $\checkmark$ \\
    \hline
    EE-7.3k-ConsensusConfig                                   & $\checkmark$ &              &          & $\checkmark$ & $\times$          & $\checkmark$  \\
    \hline
    EE-8.1a-Enode                                                    & $\checkmark$ &              &          &  $\checkmark$ & $\checkmark$ & Ethereum \\
    \hline
  \end{tabular}
  \caption{Enterprise Ethereum Client Specification Privacy Level C Requirements}
  \label{table_comparison1}
\end{table*}

\begin{table*}[t]
  \centering
  \begin{tabular}{|| l || c | c | c || *{6}{c |}}
    \hline
    Requirements                       & MUST or     & SHOULD or     & MAY               & Quorum          & Parity's                      & Hyperledger   \\
                                                & MUST NOT & SHOULD &                        &                        & Private                  & Fabric                 \\
                                                &                     & NOT        &                        &                        & Transactions        &                         \\
    \hline
    EE-5.1.1h-Organization       &                       & $\checkmark$ &                    & $\times$            &  $\times$                  & $\checkmark$   \\
    \hline
    EE-5.1.3a-SmartContractPermissioning  &     & $\checkmark$ &                    &  $\checkmark$ & $\checkmark$         & $\checkmark$   \\
    \hline
    EE-5.1.3b-RuntimeConfigUpdate             &     & $\checkmark$ &            & Partially Supported & $\times$                  & $\checkmark$   \\
        \hline
  \end{tabular}
  \caption{Additional Enterprise Ethereum Client Specification Privacy Level B Requirements}
  \label{table_comparison2}
\end{table*}

\begin{table*}[t]
  \centering
  \begin{tabular}{|| l || c | c | c || *{6}{c |}}
    \hline
    Requirements                       & MUST or     & SHOULD or     & MAY               & Quorum          & Parity's       & Hyperledger    \\
                                                & MUST NOT & SHOULD &                        &                        & Private                & Fabric              \\
                                                &                     & NOT        &                        &                        & Transactions       &                         \\
    \hline
    EE-6.1.2g-RestrictedMetadataRelayStore    &     & $\checkmark$ &       &  N/A               &  N/A                     & N/A                  \\
    \hline
    EE-6.1.2i-UnrestrictedRecipientMasking        &     & $\checkmark$ &       & N/A               & N/A                     & N/A                   \\
    \hline
    EE-6.1.2j-UnrestrictedSenderMasking           &     & $\checkmark$ &        & N/A              & N/A                     & N/A                   \\
    \hline
    EE-6.1.2k-UnrestrictedPayloadMaskingStored &    & $\checkmark$ &      & N/A              & N/A                     & N/A                   \\
    \hline
    EE-6.1.2r-PrivateTransactionAddParticipants  &     & $\checkmark$ &      & $\times$          & $\times$          & $\checkmark$  \\
    \hline
    EE-6.1.2s-PrivateTransactionConsensus        &     & $\checkmark$ &      &  $\times$         & $\checkmark$  & $\checkmark$  \\
                                                                            &      & This               &       &                         &                               &                       \\
                                                                            &      & requirement   &      &                         &                               &                        \\
                                                                            &      & should be.     &      &                         &                               &                        \\
                                                                            &      & a MUST        &      &                         &                               &                         \\
    \hline
  \end{tabular}
  \caption{Additional Enterprise Ethereum Client Specification Privacy Level A Requirements}
  \label{table_comparison3}
\end{table*}

\begin{table*}[t]
  \centering
  \begin{tabular}{|| l || c | c | c || *{6}{c |}}
    \hline
    Requirements                       & MUST or     & SHOULD or     & MAY               & Quorum          & Parity's                      & Hyperledger     \\
                                                & MUST NOT & SHOULD &                        &                        & Private                  & Fabric                           \\
                                                &                     & NOT        &                        &                        & Transactions        &                            \\
    \hline
    EE-4.1a-DApp                      &                   &                          & $\checkmark$ &          -            &                 -                & Ethereum    \\
    \hline
    EE-4.3b- FormalVerification &                & $\checkmark$       &                    &  $\checkmark$   &  $\checkmark$         & $\times$        \\
    \hline
    EE-5.1.1d-BlacklistNodes    &                      &                      & $\checkmark$ &          -               &                -                &          -           \\
    \hline
    EE-5.1.2b-BlacklistParticipants &                  &                       & $\checkmark$ &         -               &              -                &           -           \\
    \hline
    EE-5.1.3c-ConfigOptions          &                  &                       & $\checkmark$ & $\checkmark$ &               -                 & $\checkmark$  \\
    \hline
    EE-5.1.3d-LocalKeyManagement &               &                       & $\checkmark$ & $\checkmark$ & $\checkmark$         & $\checkmark$  \\
    \hline
    EE-5.1.3e-SecureExternalKeyGenStore  &    &                       & $\checkmark$ &         -              &             -                  & $\checkmark$  \\
    \hline
    EE-5.1.3f-HardwareSecurityModules &          &                       & $\checkmark$ &          -             &             -                   & $\checkmark$ \\
    \hline
    EE-5.2.1a-IntegrationLibraries               &                 &             & $\checkmark$ & $\checkmark$ &    $\checkmark$     & $\checkmark$ \\
    \hline
    EE-5.2.2a-EntDeployment                          &     & $\checkmark$ &                    & $\checkmark$ Kaleido & $\times$      & $\checkmark$ \\
    \hline
    EE-5.2.2b-EntFaultReporting                     &     & $\checkmark$ &                    & $\checkmark$ Kaleido & $\times$     &   $\checkmark$ \\
    \hline
    EE-5.2.2c-EntPerformanceManage           &     &                     & $\checkmark$ & $\checkmark$ Kaleido &             -    &   $\checkmark$  \\
    \hline
    EE-5.2.2d-EntSecurity                                &     & $\checkmark$ &                    & $\checkmark$ Kaleido &  $\times$   &   $\checkmark$  \\
    \hline
    EE-5.2.2e-EntHistoricalAnalysis                &     &                     & $\checkmark$ & $\checkmark$ Kaleido &             -     &  $\checkmark$  \\
    \hline
    EE-5.2.2f-EntManagementSystems          &     &                     & $\checkmark$ &   -           &         -       &            -             \\
    \hline
    EE-5.3.2a-InterChainInteraction                 &     &                     & $\checkmark$ & $\checkmark$ &            -                  &            -             \\
    \hline
    EE-5.3.3a-Oracles                                       &     & $\checkmark$ &                    & $\checkmark$ & $\checkmark$       & $\checkmark$    \\
    \hline
    EE-6.1.1a-OnChainSecurity                         &     & $\checkmark$ &                    & $\checkmark$  & $\checkmark$    &  $\checkmark$    \\
    \hline
    EE-6.1.2d-RestrictedMetadataMaskingStored &     &                 & $\checkmark$ &          -         &  N/A                        & $\times$           \\
    \hline
    EE-6.1.2e-RestrictedMetadataMaskingTransit &     &                 & $\checkmark$ &         -          & $\checkmark$        &  $\times$           \\
    \hline
    EE-6.1.2m-UnrestrictedMetadataMaskingStored &     &            & $\checkmark$ & N/A               & N/A                        &  N/A                  \\
    \hline
    EE-6.1.2n-UnrestrictedMetadataMaskingTransit &     &             & $\checkmark$ & N/A               & N/A                        & N/A                 \\
    \hline
    EE-6.1.2o-UnrestrictedPayloadRelayStore         &     &             & $\checkmark$ & N/A               & N/A                      & N/A                  \\
    \hline
    EE-6.1.2p-UnrestrictedMetadataRelayStore        &     &            & $\checkmark$ & N/A                & N/A                     & N/A                  \\
    \hline
    EE-6.1.3a-OffchainTrustedExecution               &     & $\checkmark$ &                 & $\times$       & $\times$                  &   $\times$    \\
    \hline
    EE-6.2.1a-ImprovedOnchainProcessing           &     & $\checkmark$ &                & $\checkmark$ & $\checkmark$ With caveats.   & $\checkmark$ \\
                                                                                &                         &             &       &                    & See section \ref{Parity-EE-6.2.1a-ImprovedOnchainProcessing}.  &     \\
    \hline
    EE-6.2.2a-OffchainProcessing                          &     & $\checkmark$ &                    & $\times$     & $\checkmark$          & $\checkmark$   \\
    \hline
    EE-6.2.3a-ArchivePrivateState                          &     & $\checkmark$ &                    & $\times$     & $\checkmark$         & $\checkmark$    \\
    \hline
    EE-6.2.3b-ComputePowerSizeParticipants       &     & $\checkmark$ & &\multicolumn{3}{c|}{See section \ref{EE-6.2.3b-ComputePowerSizeParticipants}} \\
    \hline
    EE-6.2.3c-RecentBlockAccessTime                  &     & $\checkmark$ & &\multicolumn{3}{c|}{See section \ref{EE-6.2.3b-ComputePowerSizeParticipants}} \\
    \hline
    EE-6.2.3d-NewGenesisBlock                            &     & $\checkmark$ &                    &  $\times$         & N/A                      & $\times$            \\
    \hline
    EE-7.1a-StoragePubEth                                    &     & $\checkmark$ &                    & $\checkmark$ & $\checkmark$      & Ethereum          \\
    \hline
    EE-7.1b-StorageForOptionalOffchain                   &     &                & $\checkmark$ &            -            &             -               &             -            \\
    \hline
    EE-7.1d-DataAccessSameParticipants             &     & $\checkmark$ &                    & $\checkmark$ & $\times$               & $\checkmark$    \\
    \hline
    EE-7.1f-FileDecentralizedStorage                     &     & $\checkmark$ &                    &  $\times$         & $\times$               & $\checkmark$    \\
    \hline
    EE-7.2b-EvmExtendedOpCodes                           &     &                & $\checkmark$ &           -            &              -               & Ethereum          \\
    \hline
    EE-7.2c-PublicStateSync                                  &     & $\checkmark$ &                    & $\times$          & $\checkmark$       & Ethereum          \\
    \hline
    EE-7.2d-PrecompiledContracts                             &     &                & $\checkmark$ &           -             &             -               & Ethereum          \\
    \hline
    EE-7.2e-TEE                                                      &     & $\checkmark$ &                    & $\times$          & $\times$               & $\times$            \\
    \hline
    EE-7.2f-TEEConfigurableEncryption                  &     & $\checkmark$ &                    & $\times$          & $\times$               & $\times$           \\
    \hline
    EE-7.2.1a-Finality                                               &     & $\checkmark$ &                    & $\checkmark$  & $\checkmark$     & $\checkmark$ With caveats.  \\
                                                                               &     &                         &                    &                         &                             & See section \ref{HyperLedgerFabric-EE-7.2.1a-Finality}.\\
    \hline
    EE-7.3a-MainNetConsensus                              &     & $\checkmark$ &                    & $\times$          & $\checkmark$      & Ethereum         \\
    \hline
    EE-7.3c-PrivateConsensusAlg                             &     & $\checkmark$ &                    & $\checkmark$  & $\checkmark$    & $\checkmark$   \\
    \hline
    EE-7.3d-MainNetConsensusAlg                          &     & $\checkmark$ &                    & $\times$          & $\checkmark$      & Ethereum      \\
    \hline
    EE-7.3e-SidechainConsensusAlg                            &     &                & $\checkmark$ &  $\checkmark$  & $\checkmark$    & $\checkmark$  \\
    \hline
    EE-7.3g-ConsensusAlgModularConf                  &     & $\checkmark$ &                    & $\checkmark$  &  $\times$            & $\checkmark$   \\
    \hline
    EE-7.3h-ConsensusInOutOfBand                           &     &                & $\checkmark$ & $\checkmark$  & $\checkmark$     & $\checkmark$    \\
    \hline
    EE-7.3i-ConsensusIBFT                                     &     & $\checkmark$ &                    & $\checkmark$ & $\times$               & $\times$          \\
    \hline
    EE-7.3j-ConsensusOther                                        &     &                & $\checkmark$ & $\checkmark$  & $\checkmark$     & $\checkmark$  \\
    \hline
    EE-8.1b-DevP2P                                          &     & $\checkmark$ &                           & $\times$          & $\checkmark$      & Ethereum      \\
    \hline
    EE-8.1c-Eth62Eth63                                     &     & $\checkmark$ &                           & $\times$         & $\checkmark$      & Ethereum       \\
    \hline
    EE-8.1d-NewProtocols                                           &     &                & $\checkmark$ &  $\checkmark$ &  $\checkmark$     & Ethereum       \\
    \hline
    EE-8.1e-RelayNodes                                    &     & $\checkmark$ &                           & $\times$         & $\times$               & $\times$         \\
    \hline
    EE-9a-AntiSpam                                            &     & $\checkmark$ &                           & $\times$        & Partially                & $\times$         \\
    \hline
    EE-10a-PublicEthCompatibility                      &     & $\checkmark$ &                           & $\times$       & $\checkmark$        & Ethereum        \\
    \hline
    EE-10b-ExtendedApisSuperset                              &     &                & $\checkmark$ &  $\checkmark$ & $\checkmark$      & Ethereum        \\
    \hline
    EE-11a-FastSync                                          &     & $\checkmark$ &                           & $\checkmark$ & $\checkmark$      & $\times$         \\
    \hline
    EE-11b-BackupRestore                                 &     & $\checkmark$ &                           & $\times$         & $\checkmark$       & $\times$       \\
    \hline
  \end{tabular}
  \caption{Enterprise Ethereum Client Specification Other Requirements}
  \label{table_comparison4}
\end{table*}

\begin{table*}[t]
  \centering
  \begin{tabular}{|| l || c | c | c || *{6}{c |}}
    \hline
    Requirements                              & MUST or     & SHOULD or     & MAY       & Quorum          & Parity's                      & Hyperledger     \\
                                                        & MUST NOT & SHOULD &                        &                        & Private                  & Fabric                   \\
                                                        &                     & NOT        &                        &                        & Transactions        &                            \\
    \hline
    BC-1a-ApiCallPermissioning             &            & $\checkmark$  &             & $\times$              & $\times$                 & $\checkmark$    \\
    \hline
    BC-1b-EthereumAccountWhitelist     &            & $\checkmark$  &             & $\times$             & $\times$                  & $\checkmark$   \\
    \hline
    BC1c-TransactionTypePermissioning     &            & $\checkmark$  &         & $\times$             & $\checkmark$        & $\checkmark$  \\
    \hline
    BC-1d-PrivateStateAuthenticatedEncryption & $\checkmark$ &  &             & $\times$            & $\times$                 & $\times$            \\
    \hline
    BC-2a-OrganisationallyAwareConsensus  &       & $\checkmark$ &           & $\times$            & $\times$                  &  $\times$          \\
    \hline
    BC-3a-DiscoverableBootstrapInfo           &            & $\checkmark$ &         & $\times$           & $\times$                  &   $\times$          \\
    \hline
    BC-4a-ArchitecturalDecentralization  &            & $\checkmark$  &             & $\times$           & $\checkmark$         &  $\times$           \\
    \hline
    BC-4b-PoliticalDecentralization          &            & $\checkmark$  &             & $\checkmark$  & $\checkmark$         & $\checkmark$     \\
    \hline
    BC-5a-OffchainOrgToOrg                   &            & $\checkmark$  &             & $\checkmark$  & $\checkmark$         & $\times$         \\
    \hline
    BC-5b-OffchainAll                               &            & $\checkmark$  &             & $\checkmark$  & $\checkmark$        &  $\times$         \\
    \hline
    BC-5c-OffchianAntiSpam                   &            & $\checkmark$  &             & $\checkmark$ & $\checkmark$         &  $\times$        \\
    \hline
    BC-5d-Whisper                                 &            & $\checkmark$  &             & $\checkmark$ & $\checkmark$          &  Ethereum    \\
        \hline
  \end{tabular}
  \caption{Additional Enterprise Ethereum Requirements}
  \label{table_comparison5}
\end{table*}

\begin{table*}[t]
  \centering
  \begin{tabular}{|| l || c | c | c || *{6}{c |}}
    \hline
    Requirements                                                & MUST or     & SHOULD or & MAY& Quorum   & Parity's                & Hyperledger      \\
                                                                          & MUST NOT & SHOULD     &       &                   & Private                 & Fabric               \\
                                                                          &                     & NOT            &      &                    & Transactions        &                          \\
    \hline
    SC-1a-EstablishmentNodesWhitelist                & $\checkmark$ &  &             & $\times$          & $\times$                   & $\times$       \\
    \hline
    SC-1b-EstablishmentNodesBlacklist      &                      &    & $\checkmark$ &         -              &            -                   &         -             \\
    \hline
    SC-2a-EstablishmentApiWhitelist                     & $\checkmark$ &  &             & $\times$          & $\times$                  & $\checkmark$ \\
    \hline
    SC-2b-EstablishmentApiBlacklist                      &   &  & $\checkmark$            &     -                 &            -                    &        -              \\
    \hline
    SC-3a-SidechainFindOrEstablishmentApi        & $\checkmark$ &  &             & $\times$          & $\times$                  & $\checkmark$ Partially  \\
    \hline
    SC-3b-SidechainIdentifier                                 & $\checkmark$ &  &             & $\times$          & $\times$                  & $\checkmark$    \\
    \hline
    SC-4a-Pinning                                                   & $\checkmark$ &  &             & $\checkmark$ Kaleido   & $\times$                 & $\times$         \\
    \hline
    SC-4b-PinningParticipantShielding                   &  &$\checkmark$  &             & $\checkmark$ Kaleido       & $\times$                 & $\times$          \\
    \hline
    SC-4c-PinningTransactionRateShielding          &  & $\checkmark$ &             & $\times$          & $\times$                  & $\times$         \\
    \hline
    SC-4d-PinningContesting                                 & $\checkmark$ &  &             & $\checkmark$ Kaleido     & $\times$                  & $\times$          \\
    \hline
    SC-4e-PinningCipherTextObservers                 &  &  &   $\checkmark$          & -          & -              & -         \\
    \hline
    SC-4f-PinningConfiguration                              &  &  & $\checkmark$            & $\checkmark$ Kaleido           & $\times$                  & $\times$          \\
    \hline
    SC-4g-MultipleSidechains                                & $\checkmark$ &  &             & $\times$          & $\times$                 & $\times$           \\
    \hline
    SC-5a-DataAccessDifferentParticipants           &             & $\checkmark$ &  & $\times$         &  $\times$                & $\times$            \\
    \hline
    SC-6a-SidechainArchive                       &                      &    & $\checkmark$ &        -               &                 -              &             -            \\
        \hline
  \end{tabular}
  \caption{Additional Ethereum Private Sidechain Requirements}
  \label{table_comparison6}
\end{table*}

\ifCLASSOPTIONcompsoc
  \section*{Acknowledgments}
\else
  \section*{Acknowledgment}
\fi
This research has been undertaken whilst I have been employed by ConsenSys for use within ConsenSys and as part of my PhD studies.



\bibliographystyle{IEEEtran}
\bibliography{IEEEabrv,ref}

%
%
%

\end{document}